\documentstyle[12pt]{article}

\newcommand{\be}{\begin{equation}}
\newcommand{\ee}{\end{equation}}
\newcommand{\bea}{\begin{eqnarray}}
\newcommand{\eea}{\end{eqnarray}}
\newcommand{\nn}{\nonumber}
\newcommand{\p}[1]{(\ref{#1})}

\begin{document}
\thispagestyle{empty}
\begin{flushright}Vienna, ESI-196 (1995) \\
Dubna, JINR-E2-95-53 \\
hep-th/9502073
\end{flushright}

\vskip 0.5truecm
\begin{center}
{\large\bf On the harmonic superspace geometry of $(4,4)$ supersymmetric
sigma models with torsion}
\end{center}
\vskip 0.5truecm
\begin{center}
{\large Evgenyi A. Ivanov}
\end{center}
\vskip 0.2truecm
\begin{center}
{\it Bogoliubov Laboratory of
Theoretical Physics, JINR, 141 980 Dubna, Russia} \\
{\it eivanov@thsun1.jinr.dubna.su}
\end{center}
\vskip 0.2truecm  \nopagebreak

\begin{abstract}{\small
Starting from the dual action of
$(4,4)$ $2D$ twisted multiplets in the harmonic
superspace with two independent sets of $SU(2)$ harmonic variables,
we present its generalization which hopefully provides an off-shell
description of general $(4,4)$ supersymmetric sigma models with
torsion. Like the action of the torsionless $(4,4)$
hyper-K\"ahler sigma models in the standard harmonic superspace,
it is characterized by a number of superfield potentials.
They depend on $n$ copies of
a triple of analytic harmonic $(4,4)$ superfields. As distinct
from the hyper-K\"ahler case, the potentials
prove to be severely constrained by the self-consistency condition
which stems from the commutativity of the left and right
harmonic derivatives.
We show that for $n=1$ these constraints reduce the general action
to that of $(4,4)$ twisted multiplet,
while for $n\geq 2$ there exists a wide class of new
actions which cannot be written only via twisted
multiplets.
Their most striking feature is the nonabelian and in general
nonlinear gauge invariance which
substitutes the abelian gauge symmetry of the dual action of twisted
multiplets and ensures the correct number of physical degrees of
freedom. We show, on a simple example, that these actions describe
sigma models with non-commuting left and right complex structures
on the bosonic target. }
\end{abstract}

\setcounter{page}0
\newpage
\section{Introduction}

An interesting and important class of two-dimensional supersymmetric
sigma models consists of those with $(4,4)$ worldsheet
supersymmetry. The main reason of current interest to
them is that they can
provide non-trivial backgrounds for $d=4$ strings (see, e.g.,
\cite{4string}). Relevant bosonic
target manifolds in general possess a nontrivial torsion and
two triplets of
covariantly constant complex structures (left and right ones) which
may be mutually commuting or non-commuting \cite{{GHR},{HoPa}}.
The $(4,4)$ sigma models
which can be obtained via a direct dimensional reduction of
$N=2\;\;4D$ sigma models constitute merely a subclass in the general
variety of
$(4,4)$ $2D$ sigma models; their bosonic target manifolds are
hyper-K\"ahler
(or quaternionic-K\"ahler in the case of local supersymmetry) and so
are torsionless and possess only one set of complex structures
\cite{AlFr}.
A manifestly supersymmetric off-shell description of this
latter type of
sigma models has been given in [5 - 8] in the
harmonic $N=2\;\;4D$ (or $(4,4)\;\;2D$) superspace with one set of
harmonic variables parametrizing the $SU(2)$ automorphism group
of $N=2\;\;4D$ ($(4,4)\;\;2D$) supersymmetry \cite{{GIOS},{GIKOS}}.
Later on, an analogous
formulation with the use of the same type of harmonic superspace has
been developed for sigma models with heterotic
worldsheet $(4,0)$ supersymmetry \cite{DKS} (in these models bosonic
target manifolds in general possess a torsion). One of the basic
advantages of such off-shell formulations is that they visualize
the intrinsic geometric
features of the relevant target manifolds: the corresponding
superfield
Lagrangians turn out to coincide with (or to be directly related to) the
unconstrained potentials underlying the target geometries, while the
involved superfields are identified with coordinates of some
important subspaces of the target manifolds, the analytic subspaces.
For instance, in the
torsionless $(4,4)$ case the harmonic superfield Lagrangian
is recognized as
the hyper-K\"ahler potential [7,8]. Unconstrained off-shell
formulations
provide us with an efficient tool for the explicit computation of the
bosonic metrics (e.g., hyper-K\"ahler ones in the torsionless
$(4,4)$ case) which automatically satisfy all the restrictions placed
by extended supersymmetry \cite{{GIOShk},{DVal}}. Note that these
restrictions in their own right [2,3] give no any explicit recipe for
calculating the metrics.

Since the full automorphism group of $(4,4)$
supersymmetry in two dimensions is $SO(4)_L\times SO(4)_R$,
there arises a possibility to consider more general types of
harmonic superspaces compared to the one utilized in [5 - 12].
In \cite{ISu} A. Sutulin
and the author have constructed the $(4,4)\;\;2D$ harmonic superspace
which involves two independent sets of harmonic variables
parametrizing two commuting $SU(2)$ automorphism groups in the
left and right light-cone sectors\footnote{See also \cite{BLR}.},
$SU(2)_L$ and $SU(2)_R$
(the automorphism $SU(2)$ group of the
conventional $(4,4)$  $2D$ harmonic superspace is a diagonal
in the product $SU(2)_L\times SU(2)_R$). We have shown
how to describe in this $SU(2)\times SU(2)$ harmonic superspace
the $(4,4)$ twisted supermultiplet \cite{{GHR},{IK}} and
presented the most
general off-shell action of the latter as an integral over an analytic
subspace of this superspace. The action involves
the standard number of auxilary fields (four bosonic ones) and,
in accord
with arguments of Refs. \cite{{GHR},{RSS}}, corresponds to
a general $(4,4)$ supersymmetric sigma model with
torsion and mutually commuting sets of left and right complex
structures. A new dual form
of the action in terms of unconstrained analytic superfields with an
infinite number of auxiliary fields has been also given. An interesting
peculiarity of the dual action is the abelian gauge invariance which
ensures the on-shell equivalence of this action to the original one.
We argued that this form of the action is a good starting point to
attack the problem (as yet unsolved) of constructing a manifestly
$(4,4)$ supersymmetric off-shell description of $(4,4)$ sigma models
with
non-commuting left and right complex structures.
These models {\it cannot} be described only in terms of
$(4,4)$ twisted multiplets \cite{{GHR},{RSS}}, so one is led to
seek for such generalizations of the dual action which would not
allow an equivalent formulation via $(4,4)$ harmonic superfields
representing twisted multiplets \footnote{For other proposals
of how to describe off shell $(4,4)$ and $(2,2)$ sigma models
with non-commuting complex structures see
Refs. \cite{{BLR},{IR},{DS}}.}.

In the present paper we generalize the dual action of $(4,4)$ twisted
multiplet along these lines. As the
main result, we find a wide class
of off-shell $(4,4)$ sigma model actions with a nonabelian generalization
of the abelian gauge invariance of the dual action. They cannot
be written only through $(4,4)$ twisted superfields and,
for this reason, can be thought of as corresponding to the
aforementioned more general type of $(4,4)$ sigma models.
We explicitly demonstrate the non-commutativity of the left
and right complex structures for some
interesting particular type of the actions constructed.

Our consideration is largely based upon an analogy with the
description of torsionless $N=2\;\;4D$ ($(4,4)$ in two dimensions)
hyper-K\"ahler supersymmetric sigma models in the standard
(having one set of harmonic
variables) harmonic superspace. So we start in Sect.2 by recapitulating
salient features of this description. Then in Sect.3 we recollect
the basic facts about the $SU(2)\times SU(2)$ harmonic superspace and
off-shell description of the twisted $(4,4)$ multiplet in
its framework. In Sect.4 we discuss the dual action of the latter
which involves $n$ copies of a triple
of unconstrained analytic superfields, and construct its most general
extension, proceeding from the analogy with the general hyper-K\"ahler
$(4,4)$ sigma model. This extension includes a few superfield potentials
which, as distinct from the unconstrained potentials of the
hyper-K\"ahler $(4,4)$ action,
prove to be severely constrained by the integrability condition
coming from the commutativity of the left and right harmonic
derivatives. In Sect.5 we elaborate the $n=1$ example
(with four-dimensional bosonic target) and show that the
integrability constraint just mentioned reduces the general $n=1$ action
to that of one twisted multiplet. In Sect.6 we return
to considering the generic $n\geq 2$ action. We partially solve the
integrability constraint and find a wide variety of the actions
which in general do not admit a presentation through the twisted
$(4,4)$ superfields and so encompass sigma models
with commuting as well as non-commuting left and right
complex structures. Besides the inevitable
presence of an infinite number of auxiliary fields, one more
intriguing feature of these actions is the
nonabelian and in general nonlinear
gauge invariance which generalizes the abelian gauge symmetry of the dual
action of twisted multiplets and is necessary for ensuring
the correct number of physical fields ($4n$ bosonic and
$8n$ fermionic ones). Surprisingly, the actions constructed are
bi-harmonic analogs of the action of the so called Poisson
gauge theory \cite{nonlYM} which is
a nonlinear extension of Yang-Mills theory.
We discuss in some detail their interesting subclass, viz. direct
bi-harmonic analogs of the two-dimensional Yang-Mills action.
We compute, to the first order in fields, the relevant
bosonic metric and torsion potential and show that the left and right
complex structures on the bosonic target {\it do not commute}.

\setcounter{equation}{0}
\section{Sketch of $(4,4)$ sigma models in standard harmonic
superspace}

To make further consideration self-contained, it is instructive to
start with a brief review of the off-shell formulation of
$(4,4)$ sigma models in $(4,4)\;\;2D$ harmonic superspace which is
obtained by dimensional reduction from the standard $N=2\;\;4D$ harmonic
superspace \cite{{GIOS},{GIKOS}}.
The relevant action contains no torsion in the bosonic part; the bosonic
target space metric is necessarily hyper-K\"ahler \cite{AlFr}.

The sigma models in question are described in terms of
unconstrained analytic harmonic
superfields $q^{(+)\;M}(\zeta, u)$ $(M=1,2,...2n)$
defined on
the $(2|4)$ dimensional $(4,4)\;\;2D$ analytic harmonic superspace
(see \cite{{GIO},{BGIO}} for details and terminology).
\be  \label{harmod}
(\zeta, u) = (z^{++}, z^{--}, \theta^{(+)+},  \bar{\theta}^{(+)+},
\theta^{(+)-}, \bar{\theta}^{(+)-}, u^{(+)}_i, u^{(-)}_j)\;.
\ee
Here, the harmonic variables $u^{(\pm)i}$,
$$
u^{(+)i} u^{(-)}_{i} = 1\;,
$$
parametrize the two-sphere $S^2 \sim SU(2)/U(1)$, $SU(2)$ being the
diagonal subgroup in the product of two independent $SU(2)$ automorpism
groups (the left and right ones) of the $(4,4)\;\;2D$ Poincar\'e
superalgebra. The indices $\pm$ in the parentheses refer to the harmonic
$U(1)$ charge, other $\pm$'s are $2D$ light-cone indices.

The general action of superfields $q^{(+)\;M}$ yields in the bosonic
sector a generic sigma model
on $4n$ dimensional hyper-K\"ahler manifold. The action is given
by the following integral over the $(4,4)\;\;2D$ analytic
harmonic superspace
\be \label{quaction}
S_{q} = \int \mu^{(-4)} \{ L^{(+)}_{M}(q^{(+)}, u) D^{(+2)} q^{(+)\;M} +
L^{(+4)}(q^{(+)}, u) \}\;.
\ee
The object
\be \label{Dstand}
D^{(+2)} = \partial^{(+2)} +2i(
\theta^{(+)+} \bar{\theta}^{(+)+}\partial_{++} +
\theta^{(+)-}\bar{\theta}^{(+)-}\partial_{--})\;,\;\;
(\partial^{(+2)} = u^{(+)i}\frac{\partial}{\partial u^{(-)i}})\;,
\ee
is the analyticity-preserving harmonic derivative, $\mu^{(-4)}$ is the
analytic superspace integration measure
$$
\mu^{(-4)} = d^6\zeta \;[du] =
d^2 z d^2 \theta^{(+)+} d^2 \theta^{(+)-}\;[du].
$$

Two arbitrary
potentials in the
superfield Lagrangian in (\ref{quaction}), $L^{(+)}_M (q^{(+)\;M}, u)$
and $L^{(+4)}(q^{(+)\;M}, u)$,
encode (locally) all the information
about the relevant bosonic hyper-K\"ahler manifold. The fields
parametrizing the latter appear as the first components in the harmonic
and $\theta$ expansions
of $q^{(+)\;M}$
$$
q^{(+)\;M}(\zeta, u) = q^{iM}(z)u^{(+)}_i + ... \;.
$$
The needed number of independent real fields in $q^{iM}(z)$ (just $4n$)
comes out as a result of imposing the reality condition on
the superfields $q^{(+)\; M}$:
$\widetilde{(q^{(+)\; M})} = \Omega_{MN}q^{+\;N}$, where
$\Omega_{MN}$ is a constant skew-symmetric matrix and the generalized
involution ``$\sim $'' was defined in Refs. \cite{{GIOS},{GIKOS}}.

The quantities $L^{(+)}_{M}$ and $L^{(+4)}$ have a clear geometric
meaning: these are the
hyper-K\"ahler potentials, the basic objects of unconstrained formulation
of hyper-K\"ahler geometry given for the first time in \cite{GIOS1}. There
we started from the standard definition of this geometry as a Riemann
geometry with restricted
holonomy group (in case of $4n$ dimensional manifold it should be a
subgroup of $Sp(n)$).
We extended the original (arbitrary) hyper-K\"ahler manifold by a set of
harmonic variables which parametrize the $SU(2)$ group rotating
complex structures and
then solved the constraints on the curvature by passing to a new,
analytic basis in such a harmonic extension. The main
feature of this extension which
is visualized by passing to the analytic basis is the
existence of an analytic subspace with twice as few coordinates
compared to the manifold one started with (besides the harmonic variables
the number of which is the same). The
basic geometric objects which solve the hyper-K\"ahler constraints
are just $L^{(+)}_{M}$ and $L^{(+4)}$ living as unconstrained
functions on this analytic subspace.

The fact that the action of most general $N=2\;\;4D$ ($(4,4)$ upon the
reduction to two dimensions) supersymmetric sigma model is expressed
via $L^{(+)}_{M}$ and $L^{(+4)}$, while identifying
superfields $q^{(+)\;M}$ with coordinates of an
analytic subspace of the harmonic extension of the target hyper-K\"ahler
manifold,
and the automorphism
$SU(2)$ with the $SU(2)$ group rotating complex
structures on this manifold, makes manifest the remarkable one-to-one
correspondence between $N=2\;\;4D$ ($(4,4)\;\;2D$) supersymmetry and
hyper-K\"ahler geometry \cite{AlFr}. There exists a clear analogy with
$N=1\;\;4D$
($(2,2)\;\;2D$) sigma models: the most general off-shell superfield
Lagrangian of the
latter can be interpreted as some K\"ahler potential, with the involved
chiral superfields as the coordinates
of the associated K\"ahler manifold. This makes manifest the one-to-one
correspondence between K\"ahler geometry and $N=1\;\;4D$ ($(2,2)\;\;2D$)
supersymmetry \cite{Zum}.

It is important to point out that the superfield action (\ref{quaction})
has been written and interpreted as the most general $N=2\;\;4D$
supersymmetric sigma model action \cite{{GIO},{BGIO}} {\it prior}
to recognizing
the potentials $L^{(+)}_{M}$ and $L^{(+4)}$ as the basic objects of
hyper-K\"ahler geometry and deducing them
from the primary principles of the latter in \cite{GIOS1}. Many
characteristic features of the analytic space formulation of
this geometry can
be read off by inspecting the action (\ref{quaction}). For instance,
it is
invariant under arbitrary analytic reparametrizations of $q^{(+)\;M}$
\be   \label{qrep}
\delta q^{(+)\;M} = \Lambda^{(+)\;M}(q^{(+)}, u)\;,
\ee
provided that $L^{(+)}_{M}$ and $L^{(+4)}$ transform as
\be
\delta L^{(+)}_{M} = -L^{(+)}_{N}\frac{\partial \Lambda^{(+)\;N}}
{\partial q^{(+)\;M}}\;, \;\;\delta L^{(+4)} =
-L^{(+)}_{N} \partial^{(+2)}\Lambda^{(+)\;N}\;,
\label{lrep}
\ee
as well as under the following transformations called in \cite{BGIO} the
hyper-K\"ahler ones (because these are a direct analog of K\"ahler
transformations $K(x,\bar{x}) \Rightarrow K(x,\bar{x}) + \Lambda(x) +
\bar{\Lambda}(\bar{x})$)
\be \label{hktr}
\delta q^{(+)\;M} = 0\;, \; \delta L^{(+)}_{M} =
\frac{\partial \Lambda^{(+2)}}{\partial q^{(+)\;M}}\;,\;
\delta L^{(+4)} = \partial^{(+2)} \Lambda^{(+2)}\;,\;\;
\Lambda^{(+2)} =  \Lambda^{(+2)} (q^{(+)}, u)\;.
\ee
Here $\partial^{(+2)}$ acts only on the explicit harmonics
in the arguments of $\Lambda^{(+)\;M}$, $\Lambda^{(+2)}$.
The geometric origin of these transformations have been
fully understood later on \cite{GIOS1} within the analytic
space formulation of hyper-K\"ahler manifolds. Note that
these invariances can be used to gauge
$L^{(+)}_{M}$ into its ``flat'' part $q^{(+)\;M}$
\be \label{gauge1}
L^{(+)}_{M} = - \Omega_{MN} q^{(+)\;N}\;,
\ee
thus demonstrating that the only essential hyper-K\"ahler potential is
$L^{(+4)}$ (the sign ``$-$'' in (\ref{gauge1}) ensures the
correct sign of the kinetic term of physical bosonic fields
in the component
action).

The equation of motion for $q^{(+)\;M}$ following from (\ref{quaction})
\bea
D^{(+2)}q^{(+)\;M} &=& - H^{MN} \left( \frac{\partial L^{(+4)}}
{\partial q^{(+)\;N}} - \partial^{(+2)} L^{(+)}_{N} \right)\;,
\label{qequ} \\
H^{MN}H_{NT} &=& \delta^{M}_{T}\;,\;\;\;H_{NT} \;=\;
\frac{\partial L^{(+)}_N}{\partial q^{(+)\;T}} -
\frac{\partial L^{(+)}_T}{\partial q^{(+)\;N}} \nonumber
\eea
also has a nice geometric interpretation. Defining the target
space harmonic
derivative ${\cal D}^{(+2)}$ which acts in the target analytic subspace
spanned by the coordinates $q^{(+)\;M}, u^{(\pm)i}$
\be \label{targd}
{\cal D}^{(+2)} = \partial^{(+2)} + D^{(+2)}q^{(+)\;M} \frac{\partial}
{\partial q^{(+)\;M}} \equiv \partial^{(+2)} + E^{(+3)\;M}
\frac{\partial}{\partial q^{(+)\;M}}\;,
\ee
one observes that eq. (\ref{qequ}) is none other than the expression of
the target space analytic vielbein $E^{(+3)\;M}$ in terms of the
hyper-K\"ahler potentials \cite{GIOS1}. Moreover, in the sigma model
context it is precisely eq. (\ref{qequ})
that tells us that $D^{(+2)}q^{(+)\;M} \equiv E^{(+3)\;M}$ is
target-space
analytic and, hence, that ${\cal D}^{(+2)}$ (\ref{targd}) preserves the
target space harmonic analyticity.

Summarizing, a manifestly supersymmetric off-shell formulation of
general torsionless $(4,4)$ sigma models in terms of unconstrained
harmonic-analytic superfields allows one to independently find out the
basic elements of the analytic space geometry of the target
hyper-K\"ahler manifolds. So, one way to reveal the intrinsic geometry
of the target manifolds of torsionful $(4,4)$ sigma models is
to construct the appropriate general off-shell superfield
formulation. This will be the subject of the next Sections.

In what follows we will refer to a slightly different representation
of the
general action (\ref{quaction}). Let us split the target space world
index
$M$ as $M = (i\alpha)\;,\; i=1,2;\; \alpha = 1,2,...n$ and, using the
completeness property of harmonics
$$
u^{(+)i} u^{(-)k} - u^{(+)k}u^{(-)i} = \epsilon^{ki}\;, \;\;
(\epsilon^{12} = -\epsilon_{12} = -1)\;,
$$
equivalently re-express $q^{(+)\;M} = q^{(+)\;i\alpha}$ through the pair
of analytic superfields $\omega^{\alpha}(\zeta,u)$,
$l^{(+2)\;\alpha}(\zeta,u)$
\bea
q^{(+)\;i\alpha} &=& u^{(+)i} \omega^{\alpha} - u^{(-)i} l^{(+2)\;\alpha}
\;, \nonumber \\
\omega^{\alpha} &=& u^{(-)}_i q^{(+)\;i\alpha}\;,\;\;
l^{(+2)\;\alpha} \;=\; u^{(+)}_i q^{(+)\;i\alpha}\;. \label{omlrep}
\eea
In terms of these superfields the action (\ref{quaction}) can be
rewritten as
\be \label{omlaction}
S_{\omega,l} = \int \mu^{(-4)} \{
L^{(+2)}_{\alpha}(\omega, l, u) D^{(+2)}
\omega^{\alpha} + L_{\alpha}(\omega, l, u) D^{(+2)}l^{(+2)\;\alpha} +
\tilde{L}^{(+4)}(\omega, l, u) \}\;.
\ee
To know the precise form of the relation between the potentials in
(\ref{omlaction})
and the previous ones $L^{(+)}_{M}$, $L^{(+4)}$, as well as the
$\omega, l$ realization of groups (\ref{qrep}), (\ref{hktr}), is
of no need for our further purposes. We only note that the potentials
$L^{(+2)}_{\alpha}$, $L_{\alpha}$ are also pure gauge. They can be gauged
into their flat parts
\be  \label{gauge2}
L^{(+2)}_{\alpha} = l^{(+2)\;\alpha}\;,\;\;
L_{\alpha} = -\omega^{\alpha}\;,
\ee
where, without loss of generality, we have
chosen $\Omega_{MN} = \epsilon_{ij}\delta_{\alpha\beta}$.

Note that
in this gauge and with $\tilde{L}^{(+4)}$ displaying no dependence on
$\omega^{\alpha}$, the general action reduces to
\be \label{laction}
S_{l} = \int \mu^{(-4)} \{ -2\omega^{\alpha}D^{(+2)}l^{(+2)\;\alpha} +
\tilde{L}^{(+4)}(l, u) \}\;.
\ee
This reduced action is on-shell equivalent to the general action of
$N=2$ tensor multiplets. Indeed, varying (\ref{laction}) with respect to
$\omega^\alpha$, we arrive at the action which contains only the
$\tilde{L}^{(+4)}(l, u)$ part,
\be \label{laction1}
S_{l} = \int \mu^{(-4)} \tilde{L}^{(+4)}(l, u) \;,
\ee
with the superfield $l^{(+)\;\alpha}$ subjected to the constraint
\be \label{lconstr}
D^{(+2)}l^{(+2)\;\alpha} = 0\;.
\ee
This is just the harmonic superspace action and constraint of $N=2\;\;4D$
($(4,4)\;\;2D$) tensor multiplet \cite{GIO}. Alternatively,
one could vary (\ref{laction}) with respect to $l^{(+2)\;\alpha}$
and, expressing $l^{(+2)\;\alpha}$ from the resulting algebraic
equation as a function of $D^{(+2)}\omega^\alpha$, rewrite (\ref{laction})
through the unconstrained analytic superfields $\omega^\alpha$.
This kind of $N=2\;\;4D$ ($(4,4)\;\;2D$) duality relates to each
other two different off-shell descriptions of the same scalar
supermultiplet (4+4 components on shell): with a finite number of
auxiliary fields ($l$ representation of the action) and with an
infinite number of auxiliary fields ($\omega$ representation of the action).
Note that the passing to the $\omega$ form is possible
for the general action (\ref{omlaction}) as well, because for the
superfield $l^{(+2)\;\alpha}$ the equation of motion is always algebraic,
\be \label{eql}
l^{(+)\;\alpha} \sim D^{(+2)}\omega^{\alpha} + ...\;,
\ee
and by means of this equation $l^{(+2)\;\alpha}$ can be expressed,
at least iteratively, in terms of $\omega^\alpha$.
Actually, the $l, \omega$ and $\omega$ actions are the first and
second order forms of the same general $(4,4)$ supersymmetric
hyper-K\"ahler sigma model action.

\setcounter{equation}{0}

\section{ $SU(2)\times SU(2)$ harmonic superspace}

In our further notation we will basically follow
Ref. \cite{ISu} with minor deviations.
We start with some definitions.

The standard $(4,4)\;\;2D$ superspace is defined as
$$
{\bf S}^{(1,1|4,4)} = (x^{++},
x^{--}, \theta^{+\;i\underline{k}},
\theta^{-\;a \underline{b}}).
$$
Here $+, -$ are light-cone indices and $i, \underline{k}, a,
\underline{b}$ are doublet indices of four commuting $SU(2)$
groups which constitute the full automorphism
group $SO(4)_{L}\times SO(4)_R$ of $(4,4)\;\;2D$
Poincar\'e superalgebra. The harmonic $(4,4)$ superspace constructed
in \cite{ISu} is an extension of ${\bf S}^{(1,1|4,4)}$ by two
independent sets of harmonic variables
$u^{\pm 1}_i,\;v^{\pm 1}_a$, each parametrizing one of the
$SU(2)$ factors of $SO(4)_L$ and $SO(4)_R$, respectively
(we denote them by $SU(2)_L$ and $SU(2)_R$):
$$
{\bf HS}^{(1+2, 1+2|4,4)} = {\bf S}^{(1,1|4,4)}\otimes
(u^{\pm 1}_i, v^{\pm 1}_a)\;,\;\;\;\;
u^{1\;i}u^{-1}_i = 1, \;\;v^{1\;a}v_{a}^{-1} = 1\;.
$$
The harmonics $u$ and $v$ carry two
independent $U(1)$ charges which are assumed to be strictly conserved
(like in the standard $N=2\;\;4D$ harmonic superspace
\cite{GIOS},{GIKOS}). This requirement actually implies $u$ and $v$
to parametrize the 2-spheres $S^2_L \sim SU(2)_L/U(1)_L$ and
$S^2_R \sim SU(2)_R/U(1)_R$. All superfields given on
${\bf HS}^{(1+2, 1+2|4,4)}$
possess two definite $U(1)$ charges and, correspondingly, are
assumed to be decomposable in the double harmonic series on the
above 2-spheres.

Like in the $N=2\;\;4D$ case, the main merit of passing to the
$(4,4)$ harmonic superspace in question is the existence of an analytic
subspace in it which is closed under $(4,4)$
supersymmetry and includes half of the original odd coordinates
\be \label{anss}
{\bf AS}^{(1+2, 1+2|2,2)} = (z^{++}, z^{--}, \theta^{1,0\; \underline{i}},
\theta^{0,1\; \underline{a}}, u^{\pm 1}_i,\;v^{\pm 1}_a)
\equiv (\zeta^{\mu}, u^{\pm 1}_i,\;v^{\pm 1}_a)\;,
\ee
where
$$
\theta^{1,0\; \underline{i}} = \theta^{+\;i\underline{i}}\;u^{1}_i\;,
\theta^{0,1\; \underline{a}} = \theta^{-\;a\underline{a}}\;v^{1}_a\;,
$$
and the relation between $z^{\pm\pm}$ and $x^{\pm\pm}$ can be found
in \cite{ISu}. Superfields
given on the superspace (\ref{anss}), $\Phi^{p,q}(\zeta, u, v)$
($p$ and $q$ are values of the left and right harmonic $U(1)$ charges),
are called analytic $(4,4)$ superfields.

The analytic
superspace (\ref{anss}) is real
with respect to the
generalized involution ``$\sim $'' which is the product of
ordinary complex conjugation
and an antipodal map of the 2-spheres $SU(2)_L/U(1)_L$
and $SU(2)_R/U(1)_R$
\be
\widetilde{(\theta^{1,0\;\underline{i}})} =
\theta^{1,0}_{\underline{i}}\;,\;\;
\widetilde{(u^{\pm 1\;i})} = - u^{\pm 1}_i\;,
\ee
(and similarly for $\theta^{1\;\underline{a}},
v^{\pm 1}_a$).
The analytic superfields $\Psi^{p,q}$
can be chosen real with respect to this involution, provided
$|p+q| = 2n$
\be
\widetilde{(\Psi^{p,q})} = \Psi^{p,q}\;, |p+q| = 2n\;.
\ee

In what follows we will need the fact of existence of
two mutually commuting sets of derivatives with respect to
harmonics $u^{\pm 1\;i}$ and $v^{\pm 1\;a}$, each forming an
$SU(2)$ algebra
\bea
\partial^{\pm 2,0} &=& u^{\pm 1 \;i}\frac{\partial}
{\partial u^{\mp 1 \;i}}\;,\;\;
\partial_u^0 \;=\; u^{1\;i}\frac{\partial}{\partial u^{1\;i}} -
u^{-1\;i}\frac{\partial}{\partial u^{-1\;i}}\; \nonumber \\
\partial^{0,\pm 2} &=& v^{\pm 1\;a}\frac{\partial}
{\partial v^{\mp 1 \;a}}\;,
\; \partial_v^0 \;=\; v^{1\;a}\frac{\partial}{\partial v^{1\;a}} -
v^{-1\;a}\frac{\partial}{\partial v^{-1\;a}}\;.
\label{harmder}
\eea
The full analyticity preserving harmonic
derivatives $D^{2,0}$, $D_u^0$, $D^{0,2}$, $D_v^{0}$, when applied on
analytic superfields, are given by the expressions
\bea
D^{2,0} &=& \partial^{2,0} + i\theta^{1,0}\theta^{1,0}\partial_{++}\;,
\;\;D^{0,2} \;=\;\partial^{0,2} + i\theta^{0,1}\theta^{0,1}\partial_{--}
\nonumber \\
D_u^0 &=& \partial_u^0 + \theta^{1,0\;\underline{i}}\frac{\partial}
{\partial \theta^{1,0\;\underline{i}}}\;, \;\;
D_v^0 \;=\; \partial_v^0 + \theta^{0,1\;\underline{a}}\frac{\partial}
{\partial \theta^{0,1\;\underline{a}}}\;.
\label{ander}
\eea
The operators $D_u^0$, $D_v^0$ count the $U(1)$ charges of analytic
$(4,4)$ superfields
\be
D^0_u \Phi^{p,q}(\zeta, u, v) = p\Phi^{p,q}(\zeta, u, v)\;,\;\;
D^0_v \Phi^{p,q}(\zeta, u, v) = q \Phi^{p,q}(\zeta, u, v)\;.
\ee

The last topic of this Section will be the
harmonic superspace off-shell description
of $(4,4)$ twisted chiral multiplet. Actually, the fact that
this important
multiplet has a natural formulation in the framework of the
$(4,4)$ $SU(2)\times SU(2)$ harmonic superspace furnishes the main
motivation in favour of the relevance of the latter to $(4,4)$ sigma
models with torsion.

The multiplet in question is represented
by an analytic $(4,4)$ superfield $q^{1,1}(\zeta, u,v)$ obeying
the harmonic constraints
\be
D^{2,0}q^{1,1} = D^{0,2}q^{1,1} = 0\;.
\label{qucons}
\ee
They leave in $q^{1,1}$ $8+8$ independent components \cite{ISu},
that is precisely the off-shell field content of $(4,4)$ twisted
multiplet \cite{{GHR},{IK}}.
Notice a formal similarity of the constraints
(\ref{qucons}) to the constraint (\ref{lconstr}) defining
$N=2$ tensor multiplet in the harmonic $N=2\;\;4D$ superspace.
The crucial difference between either constraints is
that (\ref{lconstr}) implies a differential condition for a vector
component of the relevant superfield, requiring it to be
divergenceless,
while this is not the case for the constraints (\ref{qucons}). These
latter constraints are purely algebraic and express the higher
dimension components of $q^{1,1}$ through $z$-derivatives of the
physical dimension ones (they leave as independent also
four auxiliary fields which enter the $\theta$ expansion of
$q^{1,1}$ as coefficients before the monomials
$\theta^{1,0\underline{i}} \theta^{0,1\underline{a}}$).

To understand the origin of the difference between
these two types of constraints,
let us perform the reduction of the $(4,4)$ $SU(2)_L\times SU(2)_R$
harmonic
superspace to the standard $(4,4)$ $SU(2)$ one. It is accomplished
by identifying harmonic variables $u^{\pm 1\;i} = v^{\pm 1\;a}$ and,
respectively, both harmonic $U(1)$ charges. The harmonic
derivative $D^{(+2)}$ (\ref{Dstand}) is recognized as the sum of the left
and right ones
$$
D^{(+2)} = D^{2,0} + D^{0,2}\;.
$$
{}From this consideration it is already clear that there is no smooth
transition
between the constraints (\ref{qucons}) and (\ref{lconstr}).
The field content of $q^{1,1}$ also changes. While
before identifying harmonics $u$ and $v$ the matrix of
physical bosons $q^{ia}(z)$ ($q^{1,1} = q^{ia}u^{1}_iv^{1}_a
+...$) comprises 4 independent fields, after the
identification this
number is reduced to 3 (only the
symmetric part of $q^{ia}$ survives).
As a result of imposing the constraint (\ref{lconstr}) on the
reduced superfield,
the lost fourth scalar field
reappears as a solution to the divergencelessness condition for the
$2D$ vector field components multiplying the $\theta$
monomials $(\theta^{(+)+})^2$,
$(\theta^{(+)-})^2$. Note that the smooth transition between the
two superfield systems becomes possible in the
dual action of $q^{1,1}$ (see below).

Despite the essential difference between the constraints
(\ref{qucons}) and (\ref{lconstr}), invariant actions of $q^{1,1}$ look
similar to those of $l^{(+2)}$ (\ref{laction1}). The general off-shell
action of
$n$ superfields $q^{1,1\;M}$
$(M=1,2,... n)$ reads
\be
S_q = \int \mu^{-2,-2}\; L^{2,2} (q^{1,1\;M}
(\zeta, u, v),\;
u,\; v)\;.
\label{genaction}
\ee
Here
$$
\mu^{-2,-2} = d^6\zeta \;[du\;dv] =
d^2z\; d^2\theta^{1,0}\;d^2 \theta^{0,1}\;[du\;dv]
$$
is the analytic superspace integration measure.
The dimensionless analytic superfield Lagrangian
$L^{2, 2} (q^{1,1\;M}, u^{\pm 1}_i, v^{\pm 1}_a) $
bears in general an arbitrary dependence on its arguments, the only
restriction being a compatibility
with the external $U(1)$ charges $2,2$. The
free action of $q^{1,1\;M}$ is given by
\be
S_{q}^{free} \sim \int \mu^{-2,-2}\; q^{1,1\;M}\;
q^{1,1\;M}\;,
\label{free}
\ee
so for consistency we are led to assume
\be
\mbox{det} \left( \frac{\partial^2 L^{2,2}}{\partial
q^{1,1\;M}
\partial q^{1,1\;N}} \right)\;|_{q^{1,1} = 0} \neq  0\;.
\ee
For completeness, we also add the constraints
on $q^{1,1\;M} (\zeta, u, v)$
\be
D^{2,0}q^{1,1\;M} = D^{0,2}q^{1,1\;M} =0\;.
\label{constrM}
\ee

The passing to the component form of the action is straightforward
\cite{ISu}. The bosonic sigma model action consists of two parts
related to each other by $(4,4)$ supersymmetry: the metric part
and the part including the torsion potential.

As an important particular example of $q^{1,1}$ action we give
the action of $(4,4)$ extension of the group manifold $SU(2)\times U(1)$
WZNW sigma model
\be \label{wzwaction}
S_{wzw} = -\frac{1}{4\kappa ^2} \int \mu^{-2,-2} \;\hat{q}^{1,1}
\hat{q}^{(1,1)}
\left( \frac{1}{(1+X)X} -\frac{\mbox{ln}(1+X)}{X^2} \right)\;.
\label{confact}
\ee
Here
\be
\hat{q}^{1,1} = q^{1,1} - c^{1,1}\;,\;X = c^{-1,-1}\hat{q}^{1,1}\;,\;
c^{\pm 1,\pm 1} = c^{ia}u^{\pm1}_iv^{\pm1}_a \;,\;
c^{ia}c_{ia} = 2\;.
\ee
Despite the presence of an extra quartet constant $c^{ia}$ in the
analytic superfield Lagrangian, the action (\ref{wzwaction})
actually does not depend
on $c^{ia}$ \cite{ISu} as it is invariant under arbitrary
rescalings and $SU(2)\times SU(2)$ rotations of this constant.

\setcounter{equation}{0}
\section{Dual form of the $q^{1,1}$ action and its generalization}

By adding the constraints (\ref{constrM}) with the superfield
Lagrange multipliers to
the general $q^{1,1}$ action
(\ref{genaction}) one puts the latter
in the form analogous
to the tensor supermultiplet master action (\ref{laction})
\be
S_{q,\omega} = \int \mu^{-2,-2} \{
 \omega^{-1,1\;M} D^{2,0} q^{1,1\;M} +
\omega^{1,-1\;M}
D^{0,2} q^{1,1\;M} + L^{2,2} (q^{1,1}, u, v)\}\;.
\label{dualgen}
\ee
The analytic superfields $q^{1,1\;M}$, $\omega^{1,-1\;M}$,
$\omega^{-1,1\;M}$
are now unconstrained and one can vary them to get the superfield
equations of motion. Varying $\omega^{1,-1\;M}$,
$\omega^{-1,1\;M}$ yields the constraints
(\ref{constrM}) and we recover the original action
(\ref{genaction}). Alternatively, one can vary (\ref{dualgen})
with respect to $q^{1,1\;M}$, which gives rise to the equation
\be
\frac{\partial L^{2,2}}{\partial q^{1,1\;M}} =
D^{2,0} \omega^{-1,1\;M} + D^{0,2}
\omega^{1,-1\;M} \equiv
A^{1,1\;M}\;.
\label{dualeq}
\ee
This algebraic equation is a kind of Legendre transformation
expressing $q^{1,1\;M}$ as a function of
$A^{1,1\;M}$
\be
\mbox{(4.2)} \Rightarrow q^{1,1\;M} = q^{1,1\;M}
(A^{1,1}, u, v) \;.
\ee
Substituting this expression back into (\ref{dualgen}), one
arrives at the dual form of the $q^{(1,1)}$ action
\bea
S_\omega  &=& \int \mu^{-2,-2} L^{2,2}_\omega
(A^{1,1}, u,v)\;,
\nonumber \\
L^{2,2}_\omega (A^{1,1}, u,v) &\equiv &
L^{2,2} (q^{1,1\;M} (A, u,v), u,v)
- q^{1,1\;M} (A,u,v)
A^{1,1\;M}\;.
\label{dualg2}
\eea

The dual action (\ref{dualg2}) provides a new off-shell
formulation of $(4,4)$ sigma models with commuting left and
right complex structures
via {\it unconstrained} analytic $(4,4)$ superfields. The most
characteristic
feature of such formulations is the presence of infinite number of
auxiliary fields \cite{{GIOS},{GIKOS}}. Thus, in the case at hand the
physical component action for $4n$ bosons
and $8n$ fermions is restored only after eliminating an
infinite tower of auxiliary fields which come from the double harmonic
expansion of superfields $\omega^{1,-1\;N}(\zeta, u, v)\;,
\;\omega^{-1,1\;N} (\zeta, u, v)$.

To see in more detail how this occurs, let us focus on
the bosonic
degrees of freedom. The action (\ref{dualgen}) originally involves three
independent superfields $q^{1,1\;N}$, $\omega^{1,-1\;N}$,
$\omega^{-1,1\;N}$,
each including $4n$ real bosonic fields in the first term of its double
harmonic expansion (higher rank bosonic fields finally prove to be
auxiliary and we should not care about them). Varying $q^{1,1\;N}$
yields an algebraic equation (\ref{dualeq}) by
which  $q^{1,1\;N}$ is eliminated in terms of the remaining two
superfields
\be \label{eqq}
q^{1,1\;N} \sim D^{2,0}\omega^{-1,1\;N} + D^{0,2}\omega^{1,-1\;N} + ...
\ee
(cf. eq. (\ref{eql})). Thereby, the number of physical dimension
bosonic fields is
reduced from $12n$ to $8n$.
However, the number of such fields carried by two
$\omega$ superfields is still twice the number of those carried by
$q^{1,1}$ in the original formulation. So one may wonder how
the on-shell equivalence of these two off-shell formulations
is achieved. The
answer is that the equivalence is guaranteed due to the
invariance of the action (\ref{dualgen}) and its $\omega$
version (\ref{dualg2}) under the abelian gauge transformations
\be
\delta \;\omega^{1,-1\;M} = D^{2,0} \sigma^{-1,-1\;M}
\;, \; \delta \;\omega^{-1,1\;M}
= - D^{0,2} \sigma^{-1,-1\;M} \;,
\label{gauge}
\ee
with  $\sigma^{-1,-1\;M} = \sigma^{-1,-1\;M}(\zeta, u, v)$
being arbitrary analytic functions. This
gauge freedom takes away just half of the lowest superisospin
multiplets in the superfields $\omega^{1,-1\;M}$,
$\omega^{-1,1\;M}$, thus restoring the correct
physical field content of the theory. For instance, the first
components in the $\theta$ expansion of these superfields are
transformed as
\be
\delta \;\omega^{1,-1\;M}_0 (z) = \partial^{2,0}
\sigma^{-1,-1\;M} (z)\;,\;
\delta\; \omega^{-1,1\;M}_0 (z) = - \partial^{0,2}
\sigma^{-1,-1\;M} (z) \;,
\ee
and one may fix the gauge so as to entirely eliminate one set of
these fields (other gauge choices are also possible). Thus,
in contrast
to the $q^{1,1}$ superfield formulation,
where the necessary set of the physical fields is ensured by
imposing the harmonic constraints on $q^{1,1}$, the same
goal in the dual formulation is achieved thanks to the gauge
freedom (\ref{gauge}) (and after eliminating
an infinite set of auxiliary fields). This gauge invariance is the main
novel feature of the dual formulation of the $q^{1,1}$ action compared to
an analogous formulation of the $l^{(+2)}$ action in the conventional
harmonic superspace. It is a necessary ingredient of the free
action of the triple $q^{1,1\;N}$, $\omega^{1,-1\;N}$,
$\omega^{-1,1\;N}$ (corresponding to the choice
$L^{2,2} = q^{1,1\;N} q^{1,1\;N}$ in (\ref{dualgen})) and one
can expect that any reasonable generalization
to the case with interaction should enjoy this important
symmetry. Below we will see that this is indeed so, the abelian gauge
invariance getting nonabelian in general.

For what follows it will be
important to note that the gauge freedom in question reflects
the commutativity of the left and right harmonic
derivatives $D^{2,0}$ and $D^{0,2}$. Indeed, the equations of
motion which follow by varying
Lagrange multipliers $\omega^{1,-1\;M}$, $\omega^{-1,1\;M}$,
viz. the constraints (\ref{constrM}), are not entirely independent:
due to the above commutativity they obey the evident integrability
condition
\be  \label{integrc}
D^{2,0} (D^{0,2}q^{1,1\;M}) - D^{0,2} (D^{2,0}q^{1,1\;M}) = 0\;.
\ee
In the simplest case we are considering, this condition
is identically satisfied (since $L^{2,2}$ does not depend
on $\omega^{1,-1\;N}$, $\omega^{-1,1\;N}$). However, in
more general cases it puts non-trivial restrictions on the
structure of the action. Below we will see that in all examples
in which the condition (\ref{integrc}) is satisfied the relevant actions
respect gauge symmetry (\ref{gauge}) or a nonabelian extension of it.

It is to the point here to adhere to a clarifying analogy with
the abelian gauge theory in two dimensions. The harmonic derivatives
$D^{2,0}, D^{0,2}$ are
analogous to the $x$ derivatives $\partial_{\mu}, \;\mu=1,2$, two
Lagrange multipliers $\omega^{1,-1\;N}$ and $- \omega^{-1,1\;N}$
being analogs of the two-dimensional $U(1)$ gauge connection $A_\mu$
(actually, of $n$ independent copies of it),
the quantity $A^{1,1\;N}$ in (\ref{dualeq}) an analog of the
gauge field strength $F_{\mu\nu} = \partial_\mu A_\nu -
\partial_\nu A_\mu \equiv \epsilon_{\mu\nu} F$. Then the dual action
(\ref{dualgen}) is analogous to the first order form of the Maxwell
action of $A_\mu$\footnote{Just as the dual action of $(4,4)$
supersymmetric hyper-K\"ahler sigma model (\ref{omlaction}) is analogous
to the first order form of a scalar field action.}, while the constraints
(\ref{qucons}) are the precise analog of the sourceless
Maxwell equation
\be \label{analogM}
\partial^{\mu}F_{\mu\nu} = \partial^{\mu}\epsilon_{\mu\nu} F = 0\;.
\ee
The self-consistency condition (\ref{integrc}) is a counterpart of
the ``kinematical'' conservation law
\be \label{analogP}
\partial^{\nu}(\partial^{\mu}F_{\mu\nu}) = 0\;.
\ee
The conservation law (\ref{analogP}) ceases to be trivial after inserting
a matter current into the r.h.s. of (\ref{analogM}): in this case it
requires the current to be conserved as a consequence of the
equations of motion,
which imposes severe restrictions on the structure of this current and
implies the gauge symmetry of the free action to extend to the whole
action (this symmetry can get nonabelian in general). Quite similarly,
after allowing for a $\omega^{1,-1\;N}$,
$\omega^{-1,1\;N}$
dependence in $L^{2,2}$ there will appear a non-zero  ``current'' in the
r.h.s. of eqs. (\ref{constrM}) and the condition (\ref{integrc})
will become the harmonic conservation law for this current,
severely restricting the structure of the latter and, hence,
of $L^{2,2}$. In the sequel we will sometimes resort to
this analogy.

The last comment concerning transformations (\ref{gauge}) is that they
define
a {\it genuine} symmetry of the actions (\ref{dualgen}), (\ref{dualg2}),
contrary, e.g., to the transformations (\ref{qrep}), (\ref{lrep}),
(\ref{hktr}) which are a kind of equivalence redefinitions of
the involved superfields and potentials. These latter transformations
leave the relevant actions form-invariant but change the precise
structure of the potentials in them. The actions
(\ref{dualgen}), (\ref{dualg2}) also possess a restricted type of
such target space form-invariance. Later on we will present
the explicit form of the latter for a generalization of (\ref{dualgen}).

Let us turn to generalizing the action
(\ref{dualgen}). As was argued in \cite{{GHR},{RSS}},
with making use of the $(4,4)$ twisted
supermultiplet alone one may construct only the
$(4,4)$ sigma models with mutually commuting left and right complex
structures.
Then a natural way to approach the problem of
constructing off-shell $(4,4)$ superfield actions with
non-commuting structures is
to seek for such generalizations of the action (\ref{dualgen}) which
{\it do not admit the passing to a pure} $q^{1,1}$
{\it form}. The rest of the paper is devoted to deducing and studying
such generalizations.

The action (\ref{dualgen}) is an analog of the dual $l^{(+2)}$ action
(\ref{laction}), the triple of superfields $q^{(1,1)}$, $\omega^{1,-1}$,
$\omega^{1,-1}$ being an analog of the pair $l^{(+2)}, \omega$.
So one may write the most general action of this triple,
making in (\ref{dualgen}) the
substitutions like those which lead from (\ref{laction}) to the general
$l,\omega$ action (\ref{omlaction}). In this way one obtains
\bea
S_{q,\omega} &=& \int \mu^{-2,-2} \{ H^{2,2} + H^{-1,1\;M}
D^{2,0}q^{1,1\;M}
+ H^{1,-1\;M}D^{0,2}q^{1,1\;M} + H^{1,1\;M}D^{0,2}\omega^{1,-1\;M}
\nonumber \\
&& + \tilde{H}^{1,1\;M}D^{2,0}\omega^{-1,1\;M}
+ H^{-1,3\;M} D^{2,0}\omega^{1,-1\;M} +
H^{3,-1\;M} D^{0,2}\omega^{-1,1\;M}  \} \nonumber \\
&\equiv & \int \mu^{-2,-2} {\cal L}^{2,2}_{q, \omega}
(q, \omega, u, v)\;,
\label{verygen}
\eea
where {\it a priori} all the potentials $H$ are arbitrary functions
of the
superfields $q^{1,1\;M}$, $\omega^{1,-1\;M}$, $\omega^{-1,1\;M}$ and
harmonics $u, v$. For the time being we leave aside the important
question of implementing the gauge freedom (\ref{gauge}) in this
action and will try to use the set of invariances of the type
(\ref{qrep}), (\ref{hktr}) to reduce the number of independent
potentials as much as possible.

One type of such invariances of the action (\ref{verygen}) is related
to reparametrizations of the involved superfields
\bea
\delta q^{1,1\;M} &=& \Lambda^{1,1\;M} (q,\omega, u,v)\;,\;\;
\delta \omega^{1,-1\;M} \;=\; \Lambda^{1,-1\;M} (q,\omega, u,v)\;,
\nonumber \\
\delta \omega^{-1,1\;M} &=& \Lambda^{-1,1\;M} (q,\omega, u,v)\;.
\label{rep}
\eea
It is straightforward to find the transformations of the potentials
such that the action is form-invariant. Their explicit structure is
not too enlightening.

Another type of invariance is similar to the hyper-K\"ahler
one (\ref{hktr})
and is related to the freedom of adding full harmonic deriavtives
to the superfield Lagrangian in (\ref{verygen})
\be
{\cal L}^{2,2}_{q,\omega} \Rightarrow {\cal L}^{2,2}_{q,\omega}  +
D^{2,0} \Lambda^{0,2} + D^{0,2} \Lambda^{2,0}\;, \label{analhk}
\ee
$$
\Lambda^{2,0} = \Lambda^{2,0}(q,\omega,u,v) \;,\;\;
\Lambda^{0,2} = \Lambda^{0,2}(q,\omega,u,v) \;.
$$
Once again, it is easy to indicate how the potentials should transform to
generate the shifts (\ref{analhk}). It will be important for our
consideration that, assuming the existence of the flat limit
(given by the action (\ref{dualgen}) with
$L^{2,2} (q, u, v) = q^{1,1\;N}q^{1,1\;N}$), the full gauge
freedom (\ref{rep}), (\ref{analhk}) can be fixed so that
\bea
H^{-1,1\;N} &=& \alpha \omega^{-1,1\;N}\;,\;\;
H^{1,-1\;N} \;=\; \beta \omega^{1,-1\;N}\;, \nonumber \\
H^{1,1\;N} &=& (1+\beta)
q^{1,1\;N}\;,\;\; \tilde{H}^{1,1\;N} \;=\; (1+\alpha)q^{1,1\;N} +
\hat{H}^{1,1\;N}\;, \label{gaugef1}
\eea
$\alpha,\;\beta$ being arbitrary parameters. In this gauge (which is
an analog of the gauges (\ref{gauge1}), (\ref{gauge2})) the action
still contains four independent potentials,
$H^{2,2}$, $H^{-1,3\;N}$,
$H^{3,-1\;N}$ and $\hat{H}^{1,1\;N}$,
\bea
S_{q,\omega} &=&
\int \mu^{-2,-2} \{ q^{1,1\;M}D^{0,2}\omega^{1,-1\;M} +
(q^{1,1\;M} + \hat{H}^{1,1\;M})D^{2,0}\omega^{-1,1\;M} \nonumber \\
&& + H^{-1,3\;M} D^{2,0}\omega^{1,-1\;M} +
H^{3,-1\;M} D^{0,2}\omega^{-1,1\;M} + H^{2,2} \}\;, \label{verygen2}
\eea
and is invariant under the following target space gauge transformations
which are a mixture of (\ref{rep}) and (\ref{analhk})
(the unconstrained analytic
parameters $\Lambda^{2,0}, \Lambda^{0,2}$ below do not
precisely coincide with those in eq. (\ref{analhk}),
but are related to them in a simple way)
\bea
\delta \hat{H}^{1,1\;M} &=& - \Lambda^{1,1\;M} +
\frac{\partial \Lambda^{0,2}}{\partial \omega^{-1,1\;M}} +
\Lambda^{1,-1\;N}\frac{\partial H^{-1,3\;N}}{\partial \omega^{-1,1\;M}} +
\Lambda^{-1,1\;N}\frac{\partial \hat{H}^{1,1\;N}}{\partial
\omega^{-1,1\;M}} \nonumber \\
\delta H^{-1,3\;M} &=&
\frac{\partial \Lambda^{0,2}}{\partial \omega^{1,-1\;M}} +
\Lambda^{1,-1\;N}\frac{\partial H^{-1,3\;N}}{\partial \omega^{1,-1\;M}} +
\Lambda^{-1,1\;N}
\frac{\partial \hat{H}^{1,1\;N}}{\partial
\omega^{1,-1\;M}}
\nonumber \\
\delta H^{3,-1\;M} &=&
\frac{\partial \Lambda^{2,0}}{\partial \omega^{-1,1\;M}} +
\Lambda^{-1,1\;N}\frac{\partial H^{3,-1\;N}}{\partial \omega^{-1,1\;M}}
\nonumber \\
\delta H^{2,2} &=& \partial^{2,0}\Lambda^{0,2} +
\partial^{0,2}\Lambda^{2,0}
+ \Lambda^{1,-1\;N}\partial^{2,0}H^{-1,3\;N} \nonumber \\
&&+ \Lambda^{-1,1\;N}
(\partial^{2,0}\hat{H}^{1,1\;N} +\partial^{0,2}H^{3,-1\;N})
\label{remgauge} \eea
with
\bea
\Lambda^{1,1\;M} &=&
\frac{\partial \Lambda^{2,0}}{\partial \omega^{1,-1\;M}} - (B^{-1})^{FN}
\frac{\partial H^{3,-1\;N}}{\partial \omega^{1,-1\;M}}
\left\{ \frac{\partial \Lambda^{0,2}}{\partial q^{1,1\;F}} -
\frac{\partial \Lambda^{2,0}}{\partial q^{1,1\;T}}
\frac{\partial H^{-1,3\;T}}{\partial q^{1,1\;F}} \right\} \nonumber \\
\Lambda^{1,-1\;M} &=& -
\frac{\partial \Lambda^{2,0}}{\partial q^{1,1\;M}} + (B^{-1})^{FN}
\frac{\partial H^{3,-1\;N}}{\partial q^{1,1\;M}}
\left\{ \frac{\partial \Lambda^{0,2}}{\partial q^{1,1\;F}} -
\frac{\partial \Lambda^{2,0}}{\partial q^{1,1\;T}}
\frac{\partial H^{-1,3\;T}}{\partial q^{1,1\;F}} \right\} \nonumber \\
\Lambda^{-1,1\;M} &=& - (B^{-1})^{NM}
\left\{ \frac{\partial \Lambda^{0,2}}{\partial q^{1,1\;N}} -
\frac{\partial \Lambda^{2,0}}{\partial q^{1,1\;T}}
\frac{\partial H^{-1,3\;T}}{\partial q^{1,1\;N}} \right\}
\label{reppar} \\
B^{MN} &=& \delta^{MN} +
\frac{\partial \hat{H}^{1,1\;M}}{\partial
q^{1,1\;N}} - \frac{\partial H^{3,-1\;M}}{\partial q^{1,1\;F}}
\frac{\partial H^{-1,3\;F}}{\partial q^{1,1\;N}}\;,\; B^{MN}(B^{-1})^{NL}
\;=\; \delta^{ML} \nonumber
\eea
(one should add, of course, the coordinate transformations
(\ref{rep}) with the parameters (\ref{reppar})).
Note that in the case of general manifold ($M=1,2...n, n>1$)
it is impossible to gauge away any of the surviving
potentials with the help of this remaining gauge freedom, though one can
still put them in the form similar to the normal gauge of the
hyper-K\"ahler
potential $L^{(+4)}$ \cite{GIOS1}. The fact that there remain
three more potentials besides $H^{2,2}$ (which is a direct analog
of $L^{(+4)}$) is
the essential difference of the considered case with torsion from
the torsionless hyper-K\"ahler case. It is worth mentioning
that upon the reduction to the $(4,4)$ $SU(2)$
harmonic superspace the superfields $\omega^{1,-1\;N}$
and $\omega^{-1,1\;N}$ in (\ref{verygen})
are identified with each other and recognized as the
single superfield $\omega^N$, $q^{1,1\;N}
\Rightarrow l^{(+2)\;N}$,  $H^{2,2} \Rightarrow L^{(+4)}$, and the
potentials $\hat{H}^{1,1\;N}$, $H^{-1,3\;N}$, $H^{3,-1\;N}$
are combined into a shift of $l^{(+2)\;N}$. This shift can be
absorbed in an equivalence redefinition of $l^{(+2)\;N}$,
after which one recovers the $\omega, l$ action (\ref{omlaction})
of the general $(4,4)$ hyper-K\"ahler sigma model in
the ``flat'' gauge (\ref{gauge2}). Note that the potentials in
\p{verygen}, \p{verygen2} will turn out to be severely
constrained, so the reduction just mentioned actually
produces some particular class of hyper-K\"ahler $(4,4)$ actions.

As was noticed in Sect.2, the $q^{(+)}$ equation of motion (\ref{qequ})
following from the general $q^{(+)}$ action (\ref{quaction}) has a
transparent interpretation within the relevant analytic target space
geometry: it
expresses the vielbein $E^{(+3)\;M} \equiv D^{(+2)}q^{(+)\;M}$ of
the analytic target space
harmonic derivative via the unconstrained hyper-K\"ahler potentials
$L^{(+4)}$, $L^{(+)}_M$. At present we have no clear understanding
which kind of the analytic target space geometry
underlies the general off-shell $(4,4)$ action with
torsion (\ref{verygen}).
By analogy with the hyper-K\"ahler case, studying this
action, the involved objects and their equations of motion
could help to clarify this point.

We will deal with the gauge-fixed action (\ref{verygen2}).
The equations of motion following from it read
\bea
&&D^{0,2}\omega^{1,-1\;M} + D^{2,0}\omega^{-1,1\;N}\left( \delta^{NM} +
\frac{\partial \hat{H}^{1,1\;N}}{\partial q^{1,1\;M}} \right)
+ D^{0,2}\omega^{-1,1\;N}
\frac{\partial H^{3,-1\;N}}{\partial q^{1,1\;M}} \nonumber \\
&&
+ D^{2,0}\omega^{1,-1\;N}
\frac{\partial H^{-1,3\;N}}{\partial q^{1,1\;M}}
= -
\frac{\partial H^{2,2}}{\partial q^{1,1\;M}}\;, \nonumber \\
&& D^{0,2}q^{1,1\;M}
+ D^{2,0}q^{1,1\;N}
\frac{\partial H^{-1,3\;M}}{\partial q^{1,1\;N}}
+ D^{2,0}\omega^{1,-1\;N} \left(
\frac{\partial H^{-1,3\;M}}{\partial \omega^{1,-1\;N}}
-  \frac{\partial H^{-1,3\;N}}{\partial \omega^{1,-1\;M}} \right) \nn \\
&& +
D^{2,0}\omega^{-1,1\;N} \left(
\frac{\partial H^{-1,3\;M}}{\partial \omega^{-1,1\;N}}
-  \frac{\partial \hat{H}^{1,1\;N}}{\partial \omega^{1,-1\;M}}
\right) \nn \\
&& -D^{0,2}\omega^{-1,1\;N}
\frac{\partial H^{3,-1\;N}}{\partial \omega^{1,-1\;M}}
=
\frac{\partial H^{2,2}}{\partial \omega^{1,-1\;M}} -
\partial^{2,0}H^{-1,3\;M}\;,
\nonumber \\
&& D^{2,0}q^{1,1\;N} \left( \delta^{MN} +
\frac{\partial \hat{H}^{1,1\;M}}{\partial q^{1,1\;N}} \right)
+
D^{0,2}q^{1,1\;N}
\frac{\partial H^{3,-1\;M}}{\partial q^{1,1\;N}}
\nonumber \\
&&+
D^{2,0}\omega^{-1,1\;N} \left(
\frac{\partial \hat{H}^{1,1\;M}}{\partial \omega^{-1,1\;N}}
-  \frac{\partial \hat{H}^{1,1\;N}}{\partial \omega^{-1,1\;M}} \right)
+
D^{2,0}\omega^{1,-1\;N} \left(
\frac{\partial \hat{H}^{1,1\;M}}{\partial \omega^{1,-1\;N}}
-  \frac{\partial H^{-1,3\;N}}{\partial \omega^{-1,1\;M}} \right)
\nonumber \\
&&+
D^{0,2}\omega^{-1,1\;N} \left(
\frac{\partial H^{3,-1\;M}}{\partial \omega^{-1,1\;N}}
-  \frac{\partial H^{3,-1\;N}}{\partial \omega^{-1,1\;M}} \right) \nn \\
&& +D^{0,2}\omega^{1,-1\;N}
\frac{\partial H^{3,-1\;M}}{\partial \omega^{1,-1\;N}}
=
\frac{\partial H^{2,2}}{\partial \omega^{-1,1\;M}} -
\partial^{2,0}\hat{H}^{1,1\;M} - \partial^{0,2} H^{3,-1\;M}. \label{eqmo}
\eea
After expressing $D^{0,2}\omega^{1,-1\;M}$ from the first of
these equations
\bea
D^{0,2}\omega^{1,-1\;M} &=&
-\frac{\partial H^{2,2}}{\partial q^{1,1\;M}} -
\left( \delta^{NM} +
\frac{\partial \hat{H}^{1,1\;N}}{\partial q^{1,1\;M}} \right)
D^{2,0}\omega^{-1,1\;N}
\nonumber \\
&&
-  \frac{\partial H^{3,-1\;N}}{\partial q^{1,1\;M}}
D^{0,2}\omega^{-1,1\;N}
- \frac{\partial H^{-1,3\;N}}{\partial q^{1,1\;M}}
D^{2,0}\omega^{1,-1\;N}\;, \label{omeq}
\eea
the remaining two can be cast in the form
\bea
D^{0,2}q^{1,1\;M} &=&
T^{1,3\;M} + T^{0,2\;NM}\;D^{2,0}\omega^{-1,1\;N} +
T^{2,0\;NM}\;D^{0,2}\omega^{-1,1\;N} \nn \\
&& + T^{-2,4\;NM}\;D^{2,0}\omega^{1,-1\;N} \equiv J^{1,3\;M}
\label{02qeq} \\
D^{2,0}q^{1,1\;M} &=&
G^{3,1\;M} + G^{2,0\;NM}\;D^{2,0}\omega^{-1,1\;N} +
G^{4,-2\;NM}\;D^{0,2}\omega^{-1,1\;N} \nn \\
&& + G^{0,2\;NM}\;D^{2,0}\omega^{1,-1\;N} \equiv J^{3,1\;M}\;.
\label{20qeq}
\eea
Here, the coefficient functions depend only on the potentials and their
derivatives. It is easy to explicitly find these functions. We leave
this for the reader as a useful exercise.

To realize the geometric meaning of the above equations, let us compare
the case under consideration with the hyper-K\"ahler one. A direct
analog of the analytic harmonic target space
$(l^{(+2)\;\alpha}, \omega^\alpha, u^{(\pm)}_i)$ in the present case
is the manifold spanned by the coordinates
$q^{1,1\;N}$, $\omega^{1,-1\:N}$, $\omega^{-1,1\;N}$ and
the target harmonics $u^{\pm 1}_i$, $v^{\pm 1}_a$.
Let us introduce, like in the hyper-K\"ahler case (eq. (\ref{targd})),
the target space harmonic derivatives ${\cal D}^{2,0}$,
${\cal D}^{0,2}$. When acting on the analytic subspace coordinates
$u$, $v$, $q^{1,1\;N}$, $\omega^{-1,1\;N}$, $\omega^{1,-1\;N}$, they are
given by the expressions
\bea
{\cal D}^{2,0} &=& \partial^{2,0} +
E^{3,1\;M}\frac{\partial}{\partial q^{1,1\;M}} +
\tilde{E}^{1,1\;M}\frac{\partial}{\partial \omega^{-1,1\;M}} +
E^{3,-1\;M}\frac{\partial}{\partial \omega^{1,-1\;M}} \;, \nonumber \\
{\cal D}^{0,2} &=& \partial^{0,2} +
E^{1,3\;M}\frac{\partial}{\partial q^{1,1\;M}} +
E^{-1,3\;M}\frac{\partial}{\partial \omega^{-1,1\;M}} +
E^{1,1\;M}\frac{\partial}{\partial \omega^{1,-1\;M}} \;,
\label{defDD} \\
E^{3,1\;M} &\equiv& D^{2,0}q^{1,1\;M}\;,\;\;
E^{1,1\;M} \;\equiv \; D^{0,2}\omega^{1,-1\;M}\;,\;\;
E^{3,-1\;M} \;\equiv \; D^{2,0}\omega^{1,-1\;M} \nonumber \\
E^{1,3\;M}  &\equiv&  D^{0,2}q^{1,1\;M}\;,\;\;
E^{-1,3\;M} \;\equiv \; D^{0,2}\omega^{-1,1\;M}\;,\;\;
\tilde{E}^{1,1\;M}  \;\equiv \;  D^{2,0}\omega^{-1,1\;M} \;,
\label{vielbeins}
\eea
where $\partial^{2,0}$, $\partial^{0,2}$ act only on the ``target''
harmonics, i.e. those appearing explicitly in the potentials and other
geometric objects. Thus the harmonic derivatives of the involved
analytic superfields
acquire the geometric meaning of vielbeins covariantizing the
flat derivatives $\partial^{2,0}$, $\partial^{0,2}$
with respect to the analytic target space gauge group
\p{remgauge}, \p{reppar}.
When promoted to the full target space,
${\cal D}^{2,0}$, ${\cal D}^{0,2}$ can get extra pieces
containing additional partial derivatives contracted with the proper
vielbeins (e.g.,
one may expect that the full harmonic target space in
the analytic basis involves, in addition to the triple of the
analytic subspace coordinates
$q^{1,1\;N}$, $\omega^{-1,1\;N}$, $\omega^{1,-1\;N}$, one more coordinate
$l^{-1,-1\;N}$ which is represented by a general harmonic superfield).
In what follows we will never specify the complete structure of
${\cal D}^{2,0}$, ${\cal D}^{0,2}$ and simply assume that
they have the proper action on all the objects depending on
harmonics $u$ and $v$. In particular, when acting on an arbitrary
analytic harmonic $(4,4)$ superfield (it can be, e.g., a
local function of superfields $q^{1,1\;N}$,
$\omega^{-1,1\;N}$, $\omega^{1,-1\;N}$ and explicitly include
harmonics $u$ and $v$),
they always coincide with  $D^{2,0}$ and $D^{0,2}$.

Keeping in mind the definition (\ref{vielbeins}), the equations of motion
\p{eqmo} (or their equivalent form \p{omeq} - \p{20qeq}) can be
interpreted in a geometric way as the relations expressing some of the
harmonic vielbeins via the potentials
$H^{2,2}$, $H^{-1,3\;N}$,
$H^{3,-1\;N}$, $\hat{H}^{1,1\;N}$. One immediately realizes what is
the main difference
from the hyper-K\"ahler relation (\ref{qequ}). Only three
harmonic vielbeins, namely, $E^{1,3\;N}$, $E^{3,1\;N}$
and some linear combination of $\tilde{E}^{1,1\;N}$, $E^{1,1\;N}$
(in \p{omeq} we have chosen it to coincide with $E^{1,1\;N}$), are
really eliminated. Three remaining vielbeins, $\tilde{E}^{1,1\;N}$,
$E^{1,-3\;N}$ and $E^{3,-1\;N}$, are not constrained by these
equations and so should be treated at this step as some independent
quantities. One cannot even conclude that they are local functions
of the analytic target space coordinates $u$, $v$, $q^{1,1\;N}$,
$\omega^{-1,1\;N}$, $\omega^{1,-1\;N}$.

In the flat target space limit (with $H^{2,2} \sim q^{1,1\;M}q^{1,1\;M}$
and all other potentials vanishing) these superfluous vielbeins besides
$\hat{H}^{1,1}$ drop out
from the equations of motion. Then it seems natural and tempting to assume
that in the case with interaction they do not contribute as well, i.e.
the corresponding coefficients in eqs. \p{omeq} - \p{20qeq} vanish.
This is indeed so and comes about as the result of taking account of the
compatibility relations which follow from the obvious commutativity
condition
\be \label{harmconsDD}
[{\cal D}^{2,0}, {\cal D}^{0,2}] = 0 \Rightarrow
\ee
\bea
{\cal D}^{0,2} E^{3,1\;N} -  {\cal D}^{2,0} E^{1,3\;N} &=& 0
\label{cons1} \\
{\cal D}^{2,0} E^{-1,3\;N} - {\cal D}^{0,2} E^{1,1\;N} &=& 0
\label{cons2} \\
{\cal D}^{0,2} E^{3,-1\;N} -  {\cal D}^{2,0} \tilde{E}^{1,1\;N} &=& 0 \;.
\label{cons3}
\eea

These relations are identically satisfied with
the definition (\ref{vielbeins}) (from the point of view of the
target space geometry these
special expressions for the harmonic vielbeins mean that
the latter are induced as a result of passing to the analytic
basis in the target space from some central basis where the
harmonic derivatives are short, ${\cal D}^{2,0} = \partial^{2,0}$,
${\cal D}^{0,2} = \partial^{0,2}$). However, once the vielbeins
are subjected to the dynamical equations \p{omeq} - \p{20qeq}, these
relations become non-trivial consistency conditions on the potentials
$H^{2,2}$, $H^{3,1}$, $H^{1,3}$, $\hat{H}^{1,1}$. Indeed, eq. \p{cons1}
together with eqs. \p{20qeq}, \p{02qeq} implies the integrability
condition
\be
D^{2,0} J^{1,3\;M} - D^{0,2} J^{3,1\;M} = 0\;, \label{bascons}
\ee
which severely constrains the coefficient functions $T$ and
$G$ in $J^{1,3\;M}$, $J^{3,1\;M}$ and, further, the potentials through
which these functions are expressed. In Sect.6 we will show that
these constraints, being combined with the target space gauge
freedom \p{remgauge}, \p{reppar}, allows one to get rid of all the
potentials in the action \p{verygen2} except for $H^{2,2}$.
Note that two other relations
in the set \p{cons1} - \p{cons3} do not place any restrictions on
the potentials, as is seen from the structure of
eqs. \p{omeq} - \p{20qeq}.

In order to gain some experience, we will first consider the $n=1$ case.

\setcounter{equation}{0}

\section{Digression: $n=1$ example}

In this case we deal with one triple of analytic superfields
$q^{1,1}$, $\omega^{1,-1}$, $\omega^{-1,1}$ and four-dimensional
manifold of physical bosons (providing the gauge
freedom (\ref{gauge}) or some its generalization hold).
The action (\ref{verygen2})
can be further simplified because
the potentials $H^{-1,3}$, $H^{3,-1}$ become pure gauge
\be \label{gaugeone}
H^{-1,3} = H^{3,-1} = 0
\ee
\be
\Rightarrow \; S^{(1)}_{q,\omega} =
\int \mu^{-2,-2} \{  q^{1,1} D^{0,2}\omega^{1,-1} +
(q^{1,1} + \hat{H}^{1,1}) D^{2,0}\omega^{-1,1} + H^{2,2} \} \;.
\label{onegen2}
\ee
Thus, the general action of the triple $q^{1,1}$,
$\omega^{1,-1}$, $\omega^{-1,1}$ is characterized by two potentials
$H^{2,2}=H^{2,2}
(q,\omega, u,v)$ and $\hat{H}^{1,1}= \hat{H}^{1,1}(q,\omega, u,v)$ which,
before enforcing the integrability condition (\ref{bascons}),
are arbitrary functions of their arguments. The action is still invariant
under the restricted class of reparametrizations preserving the gauge
(\ref{gaugeone})
\be
\delta \hat{H}^{1,1} = - \Lambda^{1,1} +
\frac{\partial \Lambda^{0,2}}{\partial \omega^{-1,1}} +
\Lambda^{-1,1}\frac{\partial \hat{H}^{1,1}}{\partial
\omega^{-1,1}} \;,\;
\delta H^{2,2} = \partial^{2,0}\Lambda^{0,2} +
\partial^{0,2}\Lambda^{2,0}  + \Lambda^{-1,1}
\partial^{2,0}\hat{H}^{1,1} \label{transf1}
\ee
\bea
\delta q^{1,1} &\equiv & \Lambda^{1,1} =
\frac{\partial \Lambda^{2,0}}{\partial \omega^{1,-1}}\;,\;
\delta \omega^{1,-1} \;\equiv \; \Lambda^{1,-1}  \;=\; -
\frac{\partial \Lambda^{2,0}}{\partial q^{1,1}}\;, \nonumber \\
\delta \omega^{-1,1} &\equiv & \Lambda^{-1,1} \;=\; - \left( 1 +
\frac{\partial \hat{H}^{1,1}}{\partial q^{1,1}} \right)^{-1}
\frac{\partial \Lambda^{0,2}}{\partial q^{1,1}} \;\equiv\; - B^{-1}
\frac{\partial \Lambda^{0,2}}{\partial q^{1,1}} \label{transf2}
\eea
\be
\frac{\partial \Lambda^{2,0}}{\partial \omega^{-1,1}} = 0 \Rightarrow
\Lambda^{2,0} = \Lambda^{2,0} (q^{1,1}, \omega^{1,-1}, u,v)\;,\;\;
\frac{\partial \Lambda^{0,2}}{\partial \omega^{1,-1}} -
B^{-1}
\frac{\partial \hat{H}^{1,1}}{\partial \omega^{1,-1}}
\frac{\partial \Lambda^{0,2}}{\partial q^{1,1}}
= 0\;. \label{restrlamb}
\ee

The set of equations (\ref{eqmo}) is also essentially simplified
\bea
D^{0,2} \omega^{1,-1}
&=& -
\frac{\partial H^{2,2}}{\partial q^{1,1}} - D^{2,0}
\omega^{-1,1} B\;, \; \;
D^{0,2} q^{1,1}
\;=\; \frac{\partial H^{2,2}}{\partial \omega^{1,-1}} +
D^{2,0}\omega^{-1,1} \frac{\partial \hat{H}^{1,1}}{\partial
\omega^{1,-1}}\;,
\nonumber \\
D^{2,0}q^{1,1} &=&
B^{-1} \left(
\frac{\partial H^{2,2}}{\partial \omega^{-1,1}} -
\partial^{2,0}\hat{H}^{1,1} - D^{2,0}\omega^{1,-1}
\frac{\partial \hat{H}^{1,1}}{\partial \omega^{1,-1}} \right) \;.
\label{eqmo1}
\eea

To extract the consequences of the integrability
condition (\ref{bascons}),
we act on the r.h.s. of the second and third equations in (\ref{eqmo1})
by $D^{2,0}$ and $D^{0,2}$,
use once again (\ref{eqmo1}) to
eliminate $D^{0,2}q^{1,1}$, $D^{2,0}q^{1,1}$
and $D^{0,2}\omega^{1,-1}$, and finally equate the obtained
expressions. Equating,
in both sides of the resulting equality, the terms without harmonic
derivatives, as well as the
coefficients before the independent structures which are
$D^{2,0}\omega^{1,-1}$, $D^{0,2}\omega^{-1,1}$,
$D^{2,0}\omega^{-1,1}$, all possible products of them,
and $(D^{2,0})^2\omega^{1,-1}$,
$(D^{0,2})^2\omega^{-1,1}$,  $D^{2,0} D^{0,2}\omega^{-1,1}$,
$(D^{2,0})^2\omega^{-1,1}$, we arrive at the set of
constraints on the potentials $H^{2,2}$ and $\hat{H}^{1,1}$. Since
we started from the equations \p{eqmo1} which respect the residual
target space gauge freedom (5.3) - (5.5), the set of
integrability constraints is also covariant. They look
rather ugly even in the $n=1$ case, so for the time being we
do not write down all of them. We firstly consider the most simple
one following from equating to zero the coefficient before the product
$\left( D^{2,0}\omega^{1,-1} \right) \left( D^{0,2}\omega^{-1,1}
\right)$ in the $n=1$ version of \p{bascons}.

It reads
\be  \label{g20}
\frac{\partial G^{0,2}}{\partial \omega^{-1,1}} = 0\;, \;\;
\left( G^{0,2}
\equiv B^{-1} \frac{\partial \hat{H}^{1,1}}{\partial
\omega^{1,-1}} \right)
\;.
\ee
It is straightforward to check that this condition is covariant
under the {\it whole} target space gauge group
(\ref{transf1}) - (\ref{restrlamb}).
Further, $G^{0,2}$ transforms as
\be \label{transfg20}
\delta G^{0,2} = - \frac{\partial^2 \Lambda^{2,0}}{\partial \omega^{1,-1}
\partial \omega^{1,-1}} + ... \;,
\ee
where dots stand for field-dependent terms. Taking into account that
both $G^{0,2}$ and $\Lambda^{2,0}$ do not depend on $\omega^{-1,1}$ and
eq. (\ref{g20}) is covariant, we conclude  from (\ref{transfg20}) that
$G^{0,2}$ can be gauged away, thus giving rise to the
gauge-fixing condition:
\be
G^{0,2} = 0 \;\;\Rightarrow \;\;
\frac{\partial \hat{H}^{1,1}}{\partial \omega^{1,-1}} = 0 \;\;
\Rightarrow  \hat{H}^{1,1} = \hat{H}^{1,1} (q^{1,1},
\omega^{-1,1}, u,v)\;.
\label{constrn1}
\ee
The residual target space gauge freedom of \p{constrn1} is
given by the transformations (\ref{transf1}) -
(\ref{restrlamb}) with the following additional restrictions on the
parameters
\bea
\frac{\partial^2 \Lambda^{2,0}}
{\partial \omega^{1,-1} \partial \omega^{1,-1}}
&=& 0 \Rightarrow
\Lambda^{2,0} = \lambda^{2,0}(q,u,v) +
\omega^{1,-1}\lambda^{1,1}(q,u,v)\;,
\nonumber \\
\frac{\partial \Lambda^{0,2}}
{\partial \omega^{1,-1}} &=& 0 \Rightarrow  \Lambda^{0,2} =
\Lambda^{0,2} (q^{1,1}, \omega^{-1,1}, u,v)\;.
\label{addrestr} \eea
Keeping in mind that $\hat{H}^{1,1}$ does not depend on
$\omega^{1,-1}$, this gauge freedom is sufficient to
entirely gauge away $\hat{H}^{1,1}$
\be \label{red}
\hat{H}^{1,1} = 0\;.
\ee

The gauge-fixed action is still invariant under
the transformations (\ref{transf1}), (\ref{transf2}) with
\be  \label{restrlamb1}
\Lambda^{2,0} = \lambda^{2,0}(q,u,v) +
\omega^{1,-1} \lambda^{1,1}(q,u,v)\;,
\;\;
\Lambda^{0,2} = \lambda^{0,2}(q,u,v) +
\omega^{-1,1} \lambda^{1,1}(q,u,v)\;.
\ee
The set of equations (\ref{eqmo1}) becomes
\be
D^{0,2}\omega^{1,-1} + D^{2,0}\omega^{-1,1}
= -
\frac{\partial H^{2,2}}{\partial q^{1,1}}\;, \; \;
D^{0,2}q^{1,1}
= \frac{\partial H^{2,2}}{\partial \omega^{1,-1}} \;, \nonumber \\
D^{2,0}q^{1,1}
 =
\frac{\partial H^{2,2}}{\partial \omega^{-1,1}}\;. \label{eqmo2}
\ee

Now it is easy to show that the remainder of the integrability
constraints is reduced to the four conditions
\bea
\frac{\partial^2 H^{2,2}}{\partial \omega^{-1,1} \partial \omega^{-1,1}}
\;=\;
\frac{\partial^2 H^{2,2}}{\partial \omega^{1,-1} \partial \omega^{1,-1}}
\;=\;
\frac{\partial^2 H^{2,2}}{\partial \omega^{1,-1} \partial \omega^{-1,1}}
&=& 0\;, \label{consH1} \\
\left( \partial^{2,0} + \frac{\partial H^{2,2}}{\partial \omega^{-1,1}}\;
\frac{\partial}{\partial q^{1,1}} \right)
\frac{\partial H^{2,2}}{\partial \omega^{1,-1}} -
\left( \partial^{0,2} + \frac{\partial H^{2,2}}{\partial \omega^{1,-1}}\;
\frac{\partial}{\partial q^{1,1}} \right)
\frac{\partial H^{2,2}}{\partial \omega^{-1,1}} &=& 0 \;.\label{consH2}
\eea
{}From eqs. (\ref{consH1}) we find
\be  \label{H1}
H^{2,2}(u,v,q,\omega) = h^{2,2}(u,v,q) + \omega^{1,-1} h^{1,3}(u,v,q)
+ \omega^{-1,1} h^{3,1}(u,v,q)\;,
\ee
after which eq. (\ref{consH2}) can be rewritten as
\be  \label{H2}
\left( \partial^{2,0} + h^{3,1}\frac{\partial}{\partial q^{1,1}}
\right) h^{1,3}
- \left( \partial^{0,2} + h^{1,3}\frac{\partial}{\partial q^{1,1}}
\right) h^{3,1} \equiv \nabla^{2,0} h^{1,3} - \nabla^{0,2} h^{3,1} = 0\;.
\ee
The action and constraints are covariant under the transformations
(\ref{transf1}), (\ref{transf2}) with the restricted parameters
(\ref{restrlamb1})
\be
\delta h^{2,2} = \nabla^{2,0} \lambda^{0,2} + \nabla^{0,2} \lambda^{2,0}
\label{transfh22}\;,\;\;\;
\delta h^{1,3} = \nabla^{0,2} \lambda^{1,1}\;,\;\;\;
\delta h^{3,1} = \nabla^{2,0} \lambda^{1,1}\;.
\label{transfh13}
\ee
It is easy to see that the action, with taking account of the constraint
(\ref{H2}), is invariant under the following generalization of
gauge transformations (\ref{gauge})
\be
\delta \omega^{1,-1} = \left( D^{2,0} + \frac{\partial h^{3,1}}
{\partial q^{1,1}}\right) \sigma^{-1,-1}\;,\;\;
\delta \omega^{-1,1} = - \left( D^{0,2} + \frac{\partial h^{1,3}}
{\partial q^{1,1}}\right) \sigma^{-1,-1}\;,
\ee
and so propagates 4 bosonic fields like the action (\ref{dualgen}).

Despite the appearance of nonlinearities, these transformations,
like (\ref{gauge}), are abelian and this property already suggests that
the action (\ref{onegen2}) with the gauge condition (\ref{red})
is actually a reparametrization of the dual form of the $q^{1,1}$ action
(\ref{dualgen}). This is indeed so. It is easy to show
(starting with a linearized level)
that the general solution to the constraint (\ref{H2}) is given by
\be \label{hsigma}
h^{1,3} = \nabla^{0,2} \Sigma^{1,1}(u,v,q)\;,\;\;
h^{3,1} = \nabla^{2,0} \Sigma^{1,1}(u,v,q)\;,
\ee
with $\Sigma^{1,1}(u,v,q)$ being an arbitrary function
(the covariant derivatives $\nabla^{2,0}$, $\nabla^{0,2}$ commute
as a consequence of (\ref{H2})). Then we can make use of
the invariance (\ref{transfh13}) to entirely gauge away
$h^{1,3}$ and $h^{3,1}$.

Thus, after fixing gauges with respect to the target space
reparametrizations and employing the consequences of the
integrability condition (\ref{cons1}), the general $n=1$
action (\ref{onegen2}) coincides, modulo a field redefinition, with the
general dual action (\ref{dualgen}) of one self-interacting
twisted $(4,4)$
multiplet. So the relevant $(4,4)$ sigma models always admit a
formulation in terms of single twisted superfield $q^{1,1}$
(constrained by (\ref{qucons})) and, in accord with the arguments of
Refs. \cite{{GHR},{RSS}}, correspond to the case of mutually commuting
left and right complex structures on the target. In the next Section
we will show that, beginning with $n=2$, this equivalence to the action
(\ref{dualgen}) ceases to hold in general.

\setcounter{equation}{0}

\section{Back to the general case}

In solving the integrability constraint \p{bascons} for the general case
with $n >1$ we will keep to the same strategy as in the $n=1$ example.
Namely,
we eliminate the harmonic derivatives $D^{2,0}q^{1,1\;M}$,
$D^{0,2}q^{1,1\;M}$, $D^{0,2}\omega^{1,-1\;M}$ in \p{bascons}
in terms of the remaining ones with the help of equations of
motion \p{omeq} - \p{20qeq} and, after this, equate to zero the
coefficients before independent structures. In this way we get a
set of constraints on the potentials $H$ which is by construction
covariant
under the target space reparametrization group \p{remgauge}, \p{reppar}.
Some of these constraints are covariant on their own, while others are
mixed under \p{remgauge}. Instead of writing down the full set of
constraints,  we will first discuss a few selected ones and show that
they, being combined with the gauge freedom \p{remgauge}, \p{reppar},
essentially reduce the number of independent potentials. This will
allow us to present the remainder of the integrability constraints in a
concise form.

As a first step we write down the constraint following from nullifying
the coefficient before $(D^{0,2})^2\omega^{-1,1\;M}$
\be
F^{4,-2\;[M,N]} \equiv
\frac{\partial H^{3,-1\;M}}{\partial \omega^{-1,1\;N}}
+
\frac{\partial H^{3,-1\;M}}{\partial q^{1,1\;S}}
\frac{\partial H^{3,-1\;N}}{\partial \omega^{1,-1\;S}} - \left( M
\leftrightarrow N \right) = 0\;. \label{31cons}
\ee
It is not difficult to verify that this constraint is covariant
with respect to \p{remgauge}, \p{reppar}
\be
\delta  F^{4,-2\;[M,N]}  = \left(
\frac{\partial \Lambda^{-1,1\;T}}{\partial \omega^{-1,1\;M}}
+ \frac{\partial \Lambda^{-1,1\;T}}{\partial q^{1,1\;S}}
\frac{\partial H^{3,-1\;M}}{\partial \omega^{1,-1\;S}} \right)
F^{4,-2\;[T,N]} - \left(M \leftrightarrow N \right)\;.
\label{Ftran}
\ee
Then it immediately follows that $H^{3,-1\;M}$ can be completely
eliminated. Indeed, using gauge freedom  \p{remgauge}, one can
gauge away the totally symmetric parts of all the coefficients in
the Taylor expansion of $H^{3,-1}$ in $\omega^{-1,1\;N}$. The remaining
parts with mixed symmetry are zero in virtue of \p{31cons}.
Thus
\be
H^{3,-1\;M} = 0\;, \label{31zero}
\ee
and the gauge function $\Lambda^{2,0}$ in \p{remgauge}, \p{reppar} gets
restricted in the following way
\be
\frac{\partial \Lambda^{2,0}}{\partial \omega^{-1,1\;M}} = 0 \;\;
\Rightarrow  \Lambda^{2,0} = \Lambda^{2,0} (q^{1,1},
\omega^{1,-1}, u, v)\;.
\label{gauge11}
\ee

With taking account of \p{31zero}, the constraints which follow from
vanishing of the coefficients before
$D^{0,2}D^{2,0} \omega^{-1,1\;N}$,  $(D^{2,0})^2 \omega^{-1,1\;N}$
and $(D^{2,0})^2 \omega^{1,-1\;N}$ in \p{bascons} are, respectively,
of the form
\bea
F^{2,0\;[M, N]} &\equiv &
\frac{\partial \hat{H}^{1,1\;M}}{\partial \omega^{-1,1\;N}} -
\frac{\partial \hat{H}^{1,1\;N}}{\partial \omega^{-1,1\;M}} = 0
\label{11cons} \\
F^{0,2\;[M,N]} &\equiv &
\left(B^{-1}\right)^{MS} \left(
\frac{\partial \hat{H}^{1,1\;S}}{\partial \omega^{1,-1\;N}} -
\frac{\partial H^{-1,3\;N}}{\partial \omega^{-1,1\;S}} \right)
- \left( M \leftrightarrow N \right) = 0 \label{11cons2} \\
F^{-2,4\;[M,N]} &\equiv &
\frac{\partial H^{-1,3\;M}}{\partial \omega^{1,-1\;N}}
-
\frac{\partial H^{-1,3\;N}}{\partial \omega^{1,-1\;M}} = 0\;.
\label{m13cons}
\eea
We will also need the constraint which comes
from putting to zero the coefficient in front of the product
$\left( D^{2,0}\omega^{1,-1\;N} \right)
\left( D^{0,2}\omega^{-1,1\;K} \right)$
\be
\frac{\partial}{\partial \omega^{-1,1\;K}}
\left\{ \left( B^{-1} \right)^{ML} \left(
\frac{\partial \hat{H}^{1,1\;L}}{\partial \omega^{1,-1\;N}} -
\frac{\partial H^{-1,3\;N}}{\partial \omega^{-1,1\;L}} \right)
\right\} = 0\;.\label{gng1} \ee
(this is the $n>1$ analog of the condition \p{g20}).

The constraint \p{m13cons} together with the gauge freedom associated
with the parameter $\Lambda^{0,2}$ (still unrestricted) allow one to
fully eliminate $H^{-1,3\;M}$
\be
H^{-1,3\;M} = 0\;. \label{13zero}
\ee
Since the expression in the curly brackets in \p{gng1} does not depend on
$\omega^{-1,1\;M}$, and its transformation law starts
with the symmetric inhomogeneous term
$$
- \frac{\partial^2 \Lambda^{2,0}}{\partial \omega^{1,-1\;M}
\partial \omega^{1,-1\;N}}\;,
$$
the part of this expression which is symmetric in the indices $M,N$
can be gauged away. Then
the constraint \p{11cons2} requires the antisymmetric part also
to vanish, whence
\be
\frac{\partial \hat{H}^{1,1\;M}}{\partial \omega^{1,-1\;N}} = 0\;.
\ee

Finally, since $\hat{H}^{1,1\;M}$ does not depend on $\omega^{1,-1\;N}$,
the residual target space gauge freedom supplemented with the
constraint \p{11cons} is still capable to completely
gauge away $\hat{H}^{1,1\;M}$
\be
\hat{H}^{1,1\;M} = 0\;.  \label{11zero}
\ee

As the result of gauge fixings \p{31zero}, \p{13zero} and \p{11zero},
the general action \p{verygen2}, the equations of motion and the
target space gauge transformations are reduced to
\be
S_{q,\omega} =
\int \mu^{-2,-2} \{ q^{1,1\;M}D^{0,2}\omega^{1,-1\;M} +
q^{1,1\;M}D^{2,0}\omega^{-1,1\;M} + H^{2,2}(q^{1,1}, \omega^{1,-1},
\omega^{-1,1}, u, v) \}\;, \label{verygen3}
\ee
\bea
D^{2,0}\omega^{-1,1\;M} + D^{0,2}\omega^{1,-1\;M} &=& -
\frac{\partial H^{2,2} (q,\omega,u,v)}{\partial q^{1,1\;M}}\;,
\nonumber \\
D^{2,0}q^{1,1\;M} \;=\;
\frac{\partial H^{2,2} (q,\omega,u,v)}{\partial \omega^{-1,1\;M}}\;, \;\;
D^{0,2}q^{1,1\;M} &=&
\frac{\partial H^{2,2} (q,\omega,u,v)}{\partial \omega^{1,-1\;M}}\;.
\label{eqmored}
\eea
\bea
\delta H^{2,2} &=&
\partial^{2,0}\Lambda^{0,2} + \partial^{0,2}\Lambda^{2,0}
\label{residH} \\
\delta q^{1,1\;N} &=& \frac{\partial \Lambda^{2,0}}{\partial
\omega^{1,-1\;N}}\;,
\;\;
\delta \omega^{1,-1\;N} =
-\frac{\partial \Lambda^{2,0}}{\partial q^{1,1\;N}}\;, \;\;
\delta \omega^{-1,1\;N} = -
\frac{\partial \Lambda^{0,2}}{\partial q^{1,1\;N}}\;. \label{residqom}
\eea
In \p{residH}, \p{residqom}
\bea \label{gremn1}
\Lambda^{2,0} &=& \lambda^{2,0}(q,u,v) + \omega^{1,-1\;N}
\lambda^{1,1\;N}(q,u,v)\;, \nonumber \\
\Lambda^{0,2} &=& \lambda^{0,2}(q,u,v) + \omega^{-1,1\;N}
\lambda^{1,1\;N}(q,u,v)\;. \label{gremn}
\eea

We stress that on the way from the most general action \p{verygen} to
the action \p{verygen3} we did not make any extra assumptions: we
only exploited the target space gauge freedom and some consequences
of the general integrability condition \p{bascons}. As we see,
the target vielbeins $E^{-1,3\;N} \equiv
D^{0,2}\omega^{-1,1\;N}$, $E^{3,-1\;N} \equiv D^{2,0}\omega^{1,-1\;N}$
entirely drop out from the dynamical equations \p{eqmored} and so
must be regarded as a
sort of auxiliary, redundant quantities in the analytic target
space geometry, in accord with the conjecture in the end of Sect.4
\footnote{One can view the conditions \p{cons2}, \p{cons3} as
the {\it definition} of $E^{-1,3\;N}$, $E^{3,-1\;N}$. These harmonic
differential equations can be solved for $E^{-1,3\;N}$, $E^{3,-1\;N}$,
thus expressing the latter through the remaining vielbeins
(nonlocally in harmonics).}. Nevertheless, one is still left with three
equations for the four unknowns
$E^{1,3\;N} \equiv D^{0,2}q^{1,1\;N}$, $E^{3,1\;N} \equiv
D^{2,0}q^{1,1\;N}$,
$E^{1,1\;N} \equiv D^{0,2}\omega^{1,-1\;N}$,
$\tilde{E}^{1,1\;N} \equiv D^{2,0}\omega^{-1,1\;N}$. As we will see soon,
the purely bosonic $\theta$ zero part of one of these
vielbeins is actually a gauge degree of freedom
due to the existence of gauge invariance generalizing the
invariance \p{gauge} of the dual twisted multiplets action \p{dualgen}.

To reveal this invariance, one should further explore the consequences
of the integrability condition \p{bascons} for the surviving potential
$H^{2,2}$.

We proceed in the same way as in the $n=1$ example.
The $n\geq 2$ generalization of the conditions
(\ref{consH1}), (\ref{consH2}) proves to be
\bea
&& \frac{\partial^2 H^{2,2}}{\partial \omega^{-1,1\;N} \partial
\omega^{-1,1\;M}}
\;=\;
\frac{\partial^2 H^{2,2}}{\partial \omega^{1,-1\;N} \partial
\omega^{1,-1\;M}}
\;=\; \frac{\partial^2 H^{2,2}}{\partial \omega^{1,-1\;(N}
\partial \omega^{-1,1\;M)}}
\;=\; 0\;, \label{consN1} \\
&& \left( \partial^{2,0} + \frac{\partial H^{2,2}}{\partial
\omega^{-1,1\;N}}
\;\frac{\partial}{\partial q^{1,1\;N}}
-{1\over 2}\; \frac{\partial H^{2,2}}{\partial q^{1,1\;N}}
\;\frac{\partial}{\partial \omega^{-1,1\;N}}
\right)
\frac{\partial H^{2,2}}{\partial \omega^{1,-1\;M}} \nonumber \\
&& -\left( \partial^{0,2} + \frac{\partial H^{2,2}}{\partial
\omega^{1,-1\;N}}
\;\frac{\partial}{\partial q^{1,1\;N}}
-{1\over 2}\; \frac{\partial H^{2,2}}{\partial q^{1,1\;N}}
\;\frac{\partial}{\partial \omega^{1,-1\;N}}
\right)
\frac{\partial H^{2,2}}{\partial \omega^{-1,1\;M}} \;=\; 0
\label{consN2}
\eea
Eqs. (\ref{consN1}) imply
\bea
H^{2,2} &=& h^{2,2}(q,u,v) + \omega^{1,-1\;N} h^{1,3\;N}(q,u,v)
+ \omega^{-1,1\;N} h^{3,1\;N}(q,u,v) \nonumber \\
&& + \;\omega^{-1,1\;N}\omega^{1,-1\;M}
h^{2,2\;[N,M]}(q,u,v)\;. \label{Hgen}
\eea
Notice the presence of the term bilinear in $\omega$s in the
general case. Substituting this expression into eq. (\ref{consN2}),
we finally derive four constraints on the potentials
$h^{2,2}$, $h^{1,3\;N}$, $h^{3,1\;N}$ and $h^{2,2\;[N,M]}$
\bea
&& \nabla^{2,0} h^{1,3\;N} - \nabla^{0,2} h^{3,1\;N} + h^{2,2\;[N,M]}
\;\frac{\partial h^{2,2}}{\partial q^{1,1\;M}} \;=\; 0 \label{1} \\
&& \nabla^{2,0} h^{2,2\;[N,M]} -
\frac{\partial h^{3,1\;N}}{\partial q^{1,1\;T}} \;h^{2,2\;[T,M]} +
\frac{\partial h^{3,1\;M}}{\partial q^{1,1\;T}}\; h^{2,2\;[T,N]} \;=\; 0
\label{2} \\
&& \nabla^{0,2} h^{2,2\;[N,M]} -
\frac{\partial h^{1,3\;N}}{\partial q^{1,1\;T}}\; h^{2,2\;[T,M]} +
\frac{\partial h^{1,3\;M}}{\partial q^{1,1\;T}}\; h^{2,2\;[T,N]} \;=\; 0
\label{3} \\
&& h^{2,2\;[N,T]}\;\frac{\partial h^{2,2\;[M,L]}}{\partial q^{1,1\;T}} +
h^{2,2\;[L,T]}\;\frac{\partial h^{2,2\;[N,M]}}{\partial q^{1,1\;T}} +
h^{2,2\;[M,T]}\;\frac{\partial h^{2,2\;[L,N]}}{\partial q^{1,1\;T}}
\;=\; 0 \;.
\label{4}
\eea
Here
\be
\nabla^{2,0} = \partial^{2,0} + h^{3,1\;N}\frac{\partial}{\partial
q^{1,1\;N}}
\;,\;\;
\nabla^{0,2} = \partial^{0,2} + h^{1,3\;N}\frac{\partial}{\partial
q^{1,1\;N}}
\;.
\ee

For further reference, we rewrite the action and the equations of motion
through the newly defined potentials
\bea
S_{q,\omega} &=&
\int \mu^{-2,-2} \{\; q^{1,1\;M}D^{0,2}\omega^{1,-1\;M} +
q^{1,1\;M}D^{2,0}\omega^{-1,1\;M} +  \omega^{1,-1\;M}h^{1,3\;M}
\nonumber \\
&&+ \omega^{-1,1\;M}h^{3,1\;M} + \omega^{-1,1\;M} \omega^{1,-1\;N}
\;h^{2,2\;[M,N]} + h^{2,2}\;\} \label{haction}
\eea
\bea
&&
\left( D^{2,0}\delta^{MN} + \frac{\partial h^{3,1\;N}}{\partial
q^{1,1\;M}}
\right)
\omega^{-1,1\;N} + \left( D^{0,2}
\delta^{MN} + \frac{\partial h^{1,3\;N}}{\partial q^{1,1\;M}} \right)
\omega^{1,-1\;N} \nonumber \\
&& + \omega^{-1,1\;S} \omega^{1,-1\;T}
\;\frac{\partial h^{2,2\;[S,T]}}{\partial q^{1,1\;M}}
\;=\; -
\frac{\partial h^{2,2}}{\partial q^{1,1\;M}}\;, \nonumber \\
&& D^{2,0}q^{1,1\;M} - h^{3,1\;M} + \omega^{1,-1\;N} h^{2,2\;[N,M]}
\;=\; 0
\nonumber \\
&& D^{0,2}q^{1,1\;M} - h^{1,3\;M} - \omega^{-1,1\;N}h^{2,2\;[N,M]}
\;=\; 0
\label{heqmo}
\eea

These action and equations enjoy a rich set of invariances.

One of them is the form-invariance under the
restricted target space reparametrizations
(\ref{residH}), (\ref{residqom}). They are realized on the superfields
and potentials in the following way
\bea
\delta q^{1,1\;N} &=& \lambda^{1,1\;N}\;,\;\; \delta \omega^{-1,1\;N}
\;=\;
-\frac{\partial \lambda^{0,2}}{\partial q^{1,1\;N}} -
\frac{\partial \lambda^{1,1\;M}}{\partial q^{1,1\;N}}\;
\omega^{-1,1\;M}\;,
\nonumber \\
\delta \omega^{1,-1\;N} &=&
-\frac{\partial \lambda^{2,0}}{\partial q^{1,1\;N}} -
\frac{\partial \lambda^{1,1\;M}}{\partial q^{1,1\;N}}\;
\omega^{1,-1\;M}\;,
\nonumber \\
\delta h^{2,2} &=& \nabla^{2,0} \lambda^{0,2} + \nabla^{0,2}
\lambda^{2,0}\;,
\nonumber \\
\delta h^{3,1\;M} &=& \nabla^{2,0}\lambda^{1,1\;M} + h^{2,2\;[M,N]}\;
\frac{\partial \lambda^{2,0}}{\partial q^{1,1\;N}} \nonumber \\
\delta h^{1,3\;M} &=& \nabla^{0,2}\lambda^{1,1\;M} - h^{2,2\;[M,N]}\;
\frac{\partial \lambda^{0,2}}{\partial q^{1,1\;N}} \nonumber \\
\delta h^{2,2\;[N,M]} &=&
\frac{\partial \lambda^{1,1\;N}}{\partial q^{1,1\;L}}\; h^{2,2\;[L,M]} -
\frac{\partial \lambda^{1,1\;M}}{\partial q^{1,1\;L}}\; h^{2,2\;[L,N]} \;.
\label{hrep}
\eea
It is a simple exercise to directly check the covariance of constraints
(\ref{1}) - (\ref{4}) under these reparametrizations.

More interesting is the gauge invariance inherent to the action
(\ref{haction}). It is a highly nontrivial nonabelian (and in
general nonlinear) generalization of the
gauge invariance (\ref{gauge})
\bea
\delta \omega^{1,-1\;M}  &=&
\left( D^{2,0}\delta^{MN} + \frac{\partial h^{3,1\;N}}
{\partial q^{1,1\;M}}
\right) \sigma^{-1,-1\;N} - \omega^{1,-1\;L} \;
\frac{\partial h^{2,2\;[L,N]}}{\partial q^{1,1\;M}}\;\sigma^{-1,-1\;N}\;,
\nonumber \\
\delta \omega^{-1,1\;M}  &=&
- \left( D^{0,2}\delta^{MN} + \frac{\partial h^{1,3\;N}}{\partial
q^{1,1\;M}} \right) \sigma^{-1,-1\;N} - \omega^{-1,1\;L} \;
\frac{\partial h^{2,2\;[L,N]}}{\partial q^{1,1\;M}}\;\sigma^{-1,-1\;N}\;,
\nonumber \\
\delta q^{1,1\;M} &=& \sigma^{-1,-1\;N} h^{2,2\;[N,M]}\;. \label{gaugenab}
\eea
As expected, the action is invariant only provided the integrability
conditions (\ref{1}) - (\ref{4}) are obeyed. In general, these gauge
transformations close with a field-dependent Lie bracket parameter.
Commuting two
such transformations on $q^{1,1\;N}$, and using the
cyclic constraint (\ref{4}), we find
\be
\delta_{br} q^{1,1\;M} = \sigma^{-1,-1\;N}_{br} h^{2,2\;[N,M]}\;, \;\;
\sigma^{-1,-1\;N}_{br} = -\sigma^{-1,-1\;L}_1 \sigma^{-1,-1\;T}_2
\frac{\partial h^{2,2\;[L,T]}}{\partial q^{1,1\;N}}\;.
\ee
We see that eq. (\ref{4}) ensures the nonlinear closure of the
algebra of gauge transformations (\ref{gaugenab}) and so it is a group
condition similar to the Jacobi identity. It is curious that the
gauge transformations (\ref{gaugenab}) with the relation (\ref{4})
are precise bi-harmonic counterparts of the basic relations of
a two-dimensional version of the recently proposed nonlinear extension of
Yang-Mills theory, so called ``Poisson gauge theory'' \cite{nonlYM}
(with the evident correspondence
$D^{2,0}, D^{0,2} \leftrightarrow \partial_\mu$;
$\omega^{1,-1\;M}, - \omega^{-1,1\;M} \leftrightarrow  A_\mu^M $).

We point out that it is the presence of the antisymmetric
potential $h^{2,2\;[N,M]}$ that makes the considered case nontrivial and,
in particular, the gauge invariance (\ref{gaugenab}) nonabelian. If
$h^{2,2\;[N,M]}$ is vanishing, the
invariance gets abelian and the constraints (\ref{1}) - (\ref{4}) except
for (\ref{1}) are identically satisfied, while (\ref{1}) can be solved
on the pattern of the $n=1$ case, eqs. (\ref{hsigma}). As a result, the
potentials $h^{1,3\;N}$, $h^{3,1\;N}$ can be gauged away using
the $\lambda^{1,1\;N}$ freedom (\ref{hrep}), and we return to the
general twisted multiplet action (\ref{dualgen}). On the contrary, with
nonvanishing $h^{2,2\;[N,M]}$ eq. (\ref{1}) does not imply
$h^{1,3\;N}$, $h^{3,1\;N}$ to be pure gauge. We cannot remove the $\omega$
dependence from second and third of eqs. (\ref{heqmo}) by any
local field redefinition with preserving harmonic analyticity. Moreover,
in contradistinction to the constraints (\ref{qucons}), these
equations are compatible
only with using the first equation. So,
the obtained system definitely does not admit in general an equivalent
description in terms of twisted $(4,4)$ analytic superfields. Hence, the
left and right complex structures on the target space can be
non-commuting and we will see soon that this is indeed so for
non-vanishing $h^{2,2\;[N,M]}$.
On the other hand, $q^{1,1\;N}$ can be expressed by
first of eqs. (\ref{heqmo}) (at least, iteratively) via
$\omega$ superfields to yield the $\omega$ representation of the action
similar to $(\ref{dualg2})$. The main distinguishing feature of this
general $\omega$ action is the nonlinear and nonabelian gauge symmetry.

To avoid a misunderstanding, we note that the analogies with
two-dimensional gauge theories are
somewhat formal because there is no any
genuine propagating gauge field among the components of the
superfields $\omega$. The only practical role of the gauge freedom
(\ref{gaugenab}) seems to consist in ensuring the correct number
of physical degrees of freedom in the action (\ref{haction})
(after elimination of $q^{1,1\;N}$ by its algebraic equation of motion).
It is also unclear, in what sense the transformations (\ref{gauge}),
(\ref{gaugenab}) could be interpreted as
gauging of some rigid ones. Indeed, in the present case the naive
definition of the rigid
group via imposing the conditions $D^{2,0}\sigma^{-1,-1\;M} =
D^{0,2}\sigma^{-1,-1\;M} =0$ leads to the trivial result
$\sigma^{-1,-1\;M} = 0$. Nevertheless, this gauge
symmetry is a necessary ingredient of the manifestly supersymmetric
off-shell unconstrained superfield description of torsionful $(4,4)$
sigma models. It should be taken into account, e.g., while quantizing
these models in the harmonic superfield formalism (one is led to
introduce the appropriate Faddeev-Popov ghosts, etc). It certainly plays
an important role in the analytic bosonic target space geometry of
the models in question. Indeed, by analogy with the
hyper-K\"ahler case \cite{GIOS1}, the basic relations of this geometry
are expected to be the $\theta$ independent parts of the superfield
equations of motion \p{eqmored} (or their more detailed form \p{heqmo}).
They relate the target space harmonic vielbeins $E^{1,3\:N}$,
$E^{3,1\:N}$, $E^{1,1\:N}$ and $\tilde{E}^{1,1\;N}$ to the potential
$H^{2,2}$. The gauge
invariance we are discussing allows us to completely gauge away one
of the
vielbeins $E^{1,1\:N}$, $\tilde{E}^{1,1\:N}$ (by gauging away
either $\omega^{1,-1\;N}|_{\theta = 0}$ or
$\omega^{-1,1\;N}|_{\theta = 0}$)
\footnote{This is not the case at the full superfield level, see eq.
\p{gaugefix}.}
and, thereby, to match the number of vielbeins with that of
independent equations. Of course, the group-theoretical and geometric
meaning of this important gauge freedom still needs to be clarified.

It remains to solve the constraints (\ref{1}) - (\ref{4}). They have
a nice geometric form and certainly encode a nontrivial geometry.
For the time being, we are
not aware of their general solution and are able to present only a
particular one. Nonetheless, it is rather interesting on its own and
seems to share most of characteristic features of the general case.

\setcounter{equation}{0}
\section{Harmonic Yang-Mills sigma models}

The particular solution we just mentioned is given by the
following ansatz
\bea
h^{1,3\;N} &=& h^{3,1\;N} \;=\; 0\;; \; h^{2,2} \;=\; h^{2,2}(t,u,v)\;,
\;
\;t^{2,2} \;=\; q^{1,1\;M}q^{1,1\;M}\;; \nonumber \\
h^{2,2\;[N,M]} &=& b^{1,1} f^{NML} q^{1,1\;L}\;, \;b^{1,1} \;=\;
b^{ia}u^1_iv^1_a\;, \; b^{ia} = \mbox{const}\;,
\label{solut}
\eea
where the constants $f^{NML}$ are real and totally antisymmetric.
The constraints
(\ref{1}) - (\ref{3}) are identically satisfied with this
ansatz, while (\ref{4}) is now none other than the Jacobi identity
which implies the constants
$f^{NML}$ to be the structure constants of some real semi-simple Lie
algebra (the minimal possibility is $n=3$, the
corresponding algebra being $so(3)$). Thus the $(4,4)$ sigma models
associated with
the ansatz \p{solut} are to be treated as a sort of Yang-Mills
theories in
the $SU(2)\times SU(2)$ harmonic superspace.
They give a natural nonabelian generalization
of the twisted multiplet sigma models with the action (\ref{dualgen})
which, as was noticed in Sect.4, are analogs of two-dimensional
abelian gauge theory. The action (\ref{haction}), equations of
motion (\ref{heqmo}) and the gauge transformation laws
(\ref{gaugenab}) specialized to the case (\ref{solut}) read
\bea
S^{YM}_{q,\omega} &=&
\int \mu^{-2,-2} \{\; q^{1,1\;M} (\; D^{0,2}\omega^{1,-1\;M} +
D^{2,0}\omega^{-1,1\;M} + b^{1,1} \;\omega^{-1,1\;L} \omega^{1,-1\;N}
f^{LNM}\; ) \nonumber \\
&&+ \;h^{2,2}(q,u,v) \} \label{haction0}
\eea
\bea
&& D^{2,0} \omega^{-1,1\;N} + D^{0,2} \omega^{1,-1\;N} +
b^{1,1}\;\omega^{-1,1\;S} \omega^{1,-1\;T} f^{STN} \;\equiv\; B^{1,1\;N}
\;=\; - \frac{\partial h^{2,2}}{\partial q^{1,1\;N}}\;, \nonumber \\
&& D^{2,0}q^{1,1\;M} + b^{1,1} \;\omega^{1,-1\;N} f^{NML}  q^{1,1\;L}
\;\equiv \; \Delta^{2,0}q^{1,1\;M}\;=\; 0
\nonumber \\
&& D^{0,2}q^{1,1\;M} - b^{1,1} \; \omega^{-1,1\;N}f^{NML}q^{1,1\;L}
\;\equiv \; \Delta^{0,2}q^{1,1\;M}\;=\; 0
\label{heqmo0} \\
&&\delta \omega^{1,-1\;M}  \;=\;
\Delta^{2,0}\sigma^{-1,-1\;M} \;, \; \delta \omega^{-1,1\;M}  \;=\;
- \Delta^{0,2}\sigma^{-1,-1\;M}\;, \nonumber \\
&& \delta q^{1,1\;M} \;=\;
b^{1,1} \;\sigma^{-1,-1\;N} f^{NML} q^{1,1\;L}\;.
\label{gaugenab0}
\eea

Now the analogy with two-dimensional nonabelian
gauge theory becomes almost literal, especially for
\be
h^{2,2} = q^{1,1\;M} q^{1,1\;M}\;. \label{free1}
\ee
With this choice,
$$
q^{1,1\;N} = -{1\over 2}\; B^{1,1\;N}
$$
in virtue of the first of eqs. (\ref{heqmo0}), then two
remaining equations are direct analogs of two-dimensional
Yang-Mills equations
\be \label{litan}
\Delta^{2,0} B^{1,1\;N} = \Delta^{0,2} B^{1,1\;N} = 0\;,
\ee
and we recognize (\ref{haction0}) and (\ref{heqmo0}) as the harmonic
counterpart of the first
order formalism of two-dimensional Yang-Mills theory.
In the general case $q^{1,1\;M}$ is a nonlinear function of
$B^{1,1\;N}$, however $B^{1,1\;N}$ still obeys the same equations
(\ref{litan}).

It is instructive to see how the fundamental integrability condition
(\ref{integrc}) is satisfied with the ansatz (\ref{solut}):
$$
[\Delta^{2,0}, \Delta^{0,2}]\; q^{1,1\;M} = - b^{1,1}\;B^{1,1\;N}
f^{NML} q^{1,1\;L} = 2b^{1,1}\; \frac{\partial h^{2,2}}{\partial
t^{2,2}}\; q^{1,1\;N}f^{NML} q^{1,1\;L} \equiv 0\;.
$$
We stress once more that in checking this condition in the nonabelian
case
one necessarily needs first of eqs. (\ref{heqmo0}), while in the abelian,
twisted multiplet case (\ref{dualgen}) the integrability condition is
satisfied without
any help from eq. (\ref{dualeq}). In other words, in the nonabelian case
we cannot interpret the $\omega$ equations of motion as some
independent kinematical constraints on $q^{1,1\;N}$: they are
self-consistent only together with the $q$ equation.
As was already mentioned, this property reflects the fact that
the class of $(4,4)$ sigma models we have found cannot be described
only in terms of twisted $(4,4)$ multiplets (of course, in general the
above gauge group has the structure of a direct product which can
include abelian factors; the
relevant $q^{1,1}$'s satisfy the linear twisted multiplet constraints
(\ref{constrM})).

An interesting feature of this ``harmonic Yang-Mills theory'' is
the presence of the doubly charged ``coupling constant''
$b^{1,1}= b^{ia}u^1_iv^1_a $,
which is necessary for the correct balance of the
harmonic $U(1)$ charges. Thus in the geometry of the considered class
of $(4,4)$ sigma models an essential role belongs to some quartet
constant $b^{ia}$. In the limit $b^{ia} \rightarrow 0$
the nonabelian structure contracts into the abelian one and we reproduce
the twisted multiplet action (\ref{dualgen}).
As we will see soon, this constant measures the ``strength'' of
non-commutativity of the left and right quaternionic structures on the
target space: in the contraction limit these structures become
mutually commuting.

In forthcoming publications we will present more detailed study of
all these issues, including those related to the
target space geometry and complex structures, at
the component level, both for the general case and the Yang-Mills
example at hand. In the rest of this Section we
give, to the first non-vanishing order in physical bosonic fields,
the bosonic metric and torsion potential, as well as the left and
right complex structures for the Yang-Mills ansatz \p{solut}.
Our purpose will be to explicitly demonstrate the non-commutativity
of complex structures for $b^{ia} \neq 0$ in \p{solut}. For simplicity
we take $h^{2,2}$ in the form \p{free1}.

As a first step we impose the following Wess-Zumino gauge with respect
to the local symmetry (\ref{gaugenab0})
\be \label{gaugefix}
\omega^{1,-1\;N} (\zeta, u, v) = \theta^{1,0\;\underline{i}} \;
\nu^{0,-1\;N}_{\underline{i}}(\zeta_R,v) + \theta^{1,0}\theta^{1,0}
\;g^{0,-1\;iN}(\zeta_R,v)u^{-1}_i
\ee
with
$$
\{ \zeta_R \} \equiv \{ x^{++},x^{--}, \theta^{0,1\;\underline{a}} \}\;.
$$
Note that with this choice there remains no any residual gauge invariance,
though all the relations below still respect a rigid invariance under
the transformations of the group with structure constants $f^{MNL}$ (it
acts as some rotations in indices $M,N, ... $).
Then we substitute (\ref{gaugefix}) into (\ref{haction0})
with $h^{2,2}$ given by (\ref{free1}), integrate over $\theta$'s and
$u$'s, eliminate infinite tails of decoupling auxiliary fields and,
finally,
obtain the physical bosons part of the action (\ref{haction0}) as the
following integral over $x$ and harmonics $v$
\be   \label{bosact}
S_{bos} = \int d^2 x [dv] \left( {i\over 2}\;g^{0,-1\;iM}(x,v)\;
\partial_{--} q^{0,1\;M}_i (x,v) \right)\;.
\ee
Here the fields $g$ and $q$ satisfy the harmonic differential
equations
\bea
&& \partial^{0,2} g^{0,-1\;iM} - 2
(b^{ka}v^{1}_a)\;f^{MNL} q^{0,1\;iN} g^{0,-1\;L}_k
\;=\; 4i \partial_{++} q^{0,1\;iM} \nonumber \\
&& \partial^{0,2} q^{0,1\;iM} - 2f^{MLN}
(b^{ka}v^1_a)\;q^{0,1\;L}_k \; q^{0,1\;iN}
\;=\; 0\;.  \label{eq12}
\eea
They are related to the initial superfields as
$$
q^{1,1\;M}(\zeta,u,v)| = q^{0,1\;iM}(x,v)u^{1}_i + ... \;, \;\;\;
g^{0,-1\;iN}(\zeta_R, v)| = g^{0,-1\;iN}(x,v)\;,
$$
where $|$ means restriction to the $\theta$ independent parts.

In order to represent the action as an integral over
$x^{++}, x^{--}$, we should solve eqs. (\ref{eq12}), substitute the
solution
into (\ref{bosact}) and perform the $v$ integration.
Here we limit ourselves to solving (\ref{eq12})
to the first
non-vanishing order in the physical bosonic  field $q^{ia\;M}(x)$,
the first component in the $v$ expansion of $q^{0,1\;iM}$
$$
q^{0,1\;iM}(x,v) = q^{ia\;M}(x)v^1_a + ... \;.
$$

Representing
(\ref{bosact}) as
\be \label{bosact1}
S_{bos} = \int d^2x  \left( G^{M\;L}_{ia\;kb} \partial_{++} q^{ia\;M}
\partial_{--} q^{kb\;L} + B^{M\;L}_{ia\;kb} \partial_{++} q^{ia\;M}
\partial_{--} q^{kb\;L} \right)
\ee
where the metric $G$ and the torsion potential $B$ are,
respectively, symmetric
and skew-symmetric with respect to the simultaneous permutation of
the left and right sets of their indices, we find
\be \label{GB}
G^{M\;L}_{ia\;kb} = \delta^{ML}\epsilon_{ik} \epsilon_{ab} -
{2\over 3} \epsilon_{ik} f^{MLN} b_{l(a} q^{l\;N}_{b)} + ...\;,
\;
B^{M\;L}_{ia\;kb}  =  {2\over 3} f^{MLN} [b_{(i a}q^N_{k)b} +
b_{(ib}q^{N}_{k)a}] + ...\;.
\ee
Note that an asymmetry between the indices $ik$ and $ab$ in the metric
is related to our choice of the WZ gauge (\ref{gaugefix}).
One could choose another gauge to restore a symmetry between
the above pairs of $SU(2)$ indices. Metrics in different
gauges are connected via the target space $q^{ia\;M}$
reparametrizations.

Finally, let us compute, again to the first order in $q^{ia\;M}$,
the relevant left and right complex structures.
According to the well-known strategy
\cite{{GHR},{HoPa},{DS}}, we need: (i) to partially go on shell by
eliminating the auxiliary
fermionic
fields; (ii) to divide four supersymmetries in every light-cone
sector into
a $N=1$ one which is realized linearly and a triplet of
nonlinearly realized extra supersymmetries; (iii) to consider the
transformation laws of the physical bosonic fields $q^{ia\;M}$ under
extra supersymmetries. The complex structures can be read off from these
transformation laws.

In our case at the step (i) we solve some harmonic differential
equations of motion in order to express an infinite tail of
auxiliary fermionic fields
in terms of the physical ones and the bosonic fields $q^{ia\;M}$.
At the step (ii) we single out the $(1,1)$ supersymmetry by
decomposing the $(4,0)$ and $(0,4)$ supersymmetry parameters
$\varepsilon^{i\underline{i}}_{-}$ and $\varepsilon^{a\underline{a}}_+$
as
$$
\varepsilon^{i\underline{i}\;+}\equiv \epsilon^{i\underline{i}}
\varepsilon^+ + i \varepsilon^{(i\underline{i})\;+} \;, \;\;
\varepsilon^{a\underline{a}\;-} \equiv \epsilon^{a\underline{a}}
\varepsilon^- + i \varepsilon^{(a\underline{a})\;-}\;,
$$
where we have kept a manifest symmetry only with respect to the diagonal
$SU(2)$ groups in the full left and right automorphism groups
$SO(4)_L$ and $SO(4)_R$.
At the step (iii) we redefine the physical fermionic fields so
that the singlet supersymmetries with the
parameters $\varepsilon_-$ and $\varepsilon_+$ are realized linearly.
We skip the details and present the final form of the on-shell
supersymmetry transformations of $q^{ia\;M}(x)$
\be
\delta q^{ia\;M} =
\varepsilon^+ \;\psi^{ia\;M}_+ +i \varepsilon^{(kj)\;+}
\;\left( F_{(kj)} \right)^{ia\;M}_{lb\;L}\;\psi^{lb\;L}_+
+
\varepsilon^- \;\chi^{ia\;M}_- + i\varepsilon^{(cd)\;-}
\;\left( \hat{F}_{(cd)} \right)^{ia\;M}_{lb\;L}\;\chi^{lb\;L}_-\;.
\ee
Introducing the matrices
$$
F^n_{(+)} \equiv  (\tau^n)^k_j F^{(j}_{\;\;\;\;k)}\;,\;\;
F^m_{(-)} \equiv  (\tau^m)^c_d \hat{F}^{(d}_{\;\;\;\;c)}\;,
$$
$\tau^n$ being Pauli matrices, we find that in the first order in
$q^{ia\;M}$ and $b^{ia}$
\bea
F^n_{(+)} &=& -i \tau^n \otimes I \otimes I + {i\over 3}\; [\;M_{(+)},
\tau^n \otimes I \otimes I\;]
\nonumber \\
F^n_{(-)} &=& -i I\otimes \tau^n \otimes I + {i\over 3}\;
[\;M_{(-)}, I\otimes \tau^n\otimes I\;] \label{cstr}
\eea
\be
\left( M_{(+)} \right)^{ia\;M}_{kb\;N} = -2\; f^{MLN}\left(
b^{(i}_{b} q^{a\;L}_{\;\;k)} + b^{(i a} q^{L}_{\;\;k)b}
\right)\;, \;\;
\left( M_{(-)} \right)^{ia\;M}_{kb\;N} =
2\;f^{MLN} \;b^i_{(b}q^{a)\;L}_k \;,\label{Matr}
\ee
where the matrix factors in the tensor products are arranged so that
they act, respectively,
on the indices $i,j,k,...$, $a,b,c,...$, $M,N,L,...$.

It is easy to see that the matrices $F^n_{(\pm)}$ to the first order
in $q$, $b$ possess all the standard properties of
complex structures needed for on-shell $(4,4)$ supersymmetry
\cite{{GHR},{HoPa}}.
In particular, they form a
quaternionic algebra
$$
F^n_{(\pm)}F^m_{(\pm)} = - \delta^{nm} + \epsilon^{nms} F^s_{(\pm)}\;,
$$
and satisfy the covariant constancy conditions
$$
{\cal D}_{lc\;K} \left( F^n_{(\pm)} \right)^{ia\;M}_{kb\;N} =
\partial_ {lc\;K}\left( F^n_{(\pm)} \right)^{ia\;M}_{kb\;N} -
\Gamma^{\;\;\;\;\;\;\;\;\;\;\;jd\;T}_{(\pm)\;lc\;K\;\;kb\;N}
\left( F^n_{(\pm)} \right)^{ia\;M}_{jd\;T} +
\Gamma^{\;\;\;\;\;\;\;\;\;\;\;ia\;M}_{(\pm)\;lc\;K\;\;jd\;T}
\left( F^n_{(\pm)} \right)^{jd\;T}_{kb\;N} = 0
$$
with
$$
\Gamma^{\;\;\;\;\;\;\;\;\;\;\;jd\;T}_{(\pm)\;lc\;M\;\;kb\;N} \equiv
\Gamma^{\;\;\;jd\;T}_{lc\;M\;\;kb\;N} \mp
T^{\;\;\;jd\;T}_{lc\;M\;\;kb\;N}\;,
$$
where $\Gamma$ is the standard Riemann connection for the
metric (\ref{GB}) and $T$ is the torsion
$$
T_{ia\;M\;\;kb\;N\;\;ld\;T} = {1\over 2} \left( \partial_{ia\;M}
B^{N\;T}_{kb\;ld} + \;\;cyclic \right)\;.
$$
It is also straightforward to check two remaining
conditions of the on-shell $(4,4)$ supersymmetry [2,3].
In the present case all these
requirements are guaranteed to be automatically fulfilled because
we proceeded from a manifestly $(4,4)$ supersymmetric off-shell
superfield formulation.

It remains to compute the commutator of complex structures.
We find (again, to the first order in fields)
\bea
[\;F^n_{(+)}, F^m_{(-)}\;] &=&
(\tau^n\otimes I\otimes I) M_{(-)}
(I\otimes \tau^m\otimes I) +
(I\otimes \tau^m\otimes I) M_{(-)}
(\tau^n\otimes I\otimes I) \nonumber \\
&-& (\tau^n\otimes
\tau^m\otimes I)
M_{(-)} -
M_{(-)}(\tau^n\otimes
\tau^m\otimes I)
 \;\neq \; 0\;.
\label{commstr}
\eea

Thus in the present case in the bosonic sector we encounter a
more general
geometry compared to the one associated with twisted $(4,4)$ multiplets.
The basic characteristic
feature of this geometry is the non-commutativity  of the left and right
complex structures.
It is easy to check this property also for general
potentials $h^{2,2}(q,u,v)$ in (\ref{haction0})\footnote{This means,
in particular, that a subclass of metrics associated with twisted
$(4,4)$ multiplets, for dimensions $4n,\;n\geq 3$, admits a deformation
which preserves $(4,4)$ SUSY but makes the left and right complex
structures non-commuting.}. It
seems obvious that the general case (\ref{haction}),
({\ref{1}) - (\ref{4}) reveals the same feature.
Stress once more that this important property
is related in a puzzling way to
the nonabelian structure of the analytic superspace
actions (\ref{haction0}),
(\ref{haction}): the coupling constant $b^{1,1}$
(or the Poisson potential
$h^{2,2\;[M,N]}$ in the general case) measures the strength of the
non-commutativity of complex structures.

The main purpose of this Section was to explicitly show that in the
$(4,4)$ models we
have constructed the left and right
complex structures on the bosonic target do not commute.
For full understanding of the geometry of these models, at least
in the particular case discussed in this Section, and for clarifying its
relation to the known examples, e.g., to the group manifold ones
\cite{Belg1}, we need the explicit form of the metrics and torsion
potentials in (\ref{bosact1}). This amounts to finding the
complete (non-iterative)
solution to eqs. \p{eq12} and their generalization to the case of
non-trivial potentials $h^{2,2}(t,u,v)$ in \p{haction0}. A work along
this line is now in progress. We wish to point out that one of the merits
of the off-shell formulation proposed consists in the fact that,
similarly to the case of hyper-K\"ahler $(4,4)$ sigma models
\cite{GIOShk} or $(4,0)$ models \cite{DVal}, we can {\it explicitly}
compute the bosonic metrics starting from the unconstrained superfield
action \p{haction0} (or its generalization corresponding to the general
solution of constraints \p{1} - \p{4}). These metrics are guaranteed to
satisfy all the conditions of on-shell $(4,4)$ supersymmetry.

Though we are not yet aware of the detailed properties of the
corresponding bosonic metrics (singularities, etc.),
in the particular case \p{solut} we know some of their isometries.
Namely, as was already mentioned, the action \p{haction0} and
its bosonic part \p{bosact1} (for any choice of $h^{2,2}(t, u, v)$
in \p{solut}) respect invariance under the global transformations of
the group with structure constants $f^{MNL}$. This suggests a
link with the group manifold $(4,4)$ models \cite{Belg1}.

Our last comment concerns the relation with
the recent paper by Delduc and Sokatchev \cite{DS}. They
studied a superfield description of $(2,2)$ sigma models with
non-commuting structures and found a set of nonlinear constraints on the
Lagrangian which somewhat resemble eqs. \p{1} - \p{4}. An
essential difference of their approach from ours seems to
lie in that it does not allow a smooth limiting transition to
the case with commuting structures.

\setcounter{equation}{0}

\section{Summary and outlook}
For reader's convenience, we summarize here the basic
steps and results of our analysis.

We proceeded from the dual action (\ref{dualgen}) of $(4,4)$
twisted multiplet in the analytic harmonic $SU(2)\times SU(2)$ superspace
and wrote down its most general conceivable extension (\ref{verygen})
involving $n$ copies of the triple of analytic harmonic superfields
$q^{1,1\;M}$, $\omega^{1,-1\;M}$, $\omega^{-1,1\;M}$ $(M=1,...n)$. Then,
using a freedom with respect to the redefinitions (\ref{rep}) and
(\ref{analhk}), we reduced it to the form (\ref{verygen2}).
It has been further simplified, to the form \p{verygen3}, by using
the residual target space gauge freedom \p{remgauge}, \p{reppar} together
with some consequences of the integrability condition \p{bascons} which
stems from the commutativity of the harmonic derivatives $D^{2,0}$ and
$D^{0,2}$. After this we studied further restrictions imposed on the
structure of the action \p{verygen3} by the integrability condition
(\ref{bascons}). The latter entirely fixes the $\omega$ dependence of
the superfield Lagrangian, bringing the
action to the form (\ref{haction}) with the potentials $h^{2,2}$,
$h^{1,3\;N}$, $h^{3,1\;N}$ and $h^{2,2\;[N,M]}$
constrained by eqs. (\ref{1}) - (\ref{4}). The action (\ref{haction})
reveals new features compared to the twisted multiplet action
(\ref{dualgen}) only provided the potential $h^{2,2\;[N,M]}$
is non-vanishing; otherwise, (\ref{haction}) can be reduced to
(\ref{dualgen}) by a field redefinition.
For $n=1$ the potential $h^{2,2\;[N,M]}$ identically vanishes, so the
novel class of $(4,4)$ sigma model actions with non-zero
$h^{2,2\;[N,M]}$ exists beginning with $n=2$. Its main novelty
is the nonabelian and
in general nonlinear gauge invariance (\ref{gaugenab}) which
substitutes the abelian gauge invariance (\ref{gauge}) of the
twisted multiplets action.
These new actions involve an infinite number of auxiliary fields and
do not admit a formulation in terms of the twisted $(4,4)$
superfields only. They provide an off-shell description of
$(4,4)$ sigma models with non-commuting left and right triplets of
complex structures.

There remains a lot of things to be done and questions to be answered.
Besides a general problem of inquiring the intrinsic geometric aspects
of the action (\ref{haction}) and constraints (\ref{1}) - (\ref{4})
as well as revealing their links with the full target space geometry,
there are a few more specific (and urgent) ones two of which we will
mention here.

An interesting problem is to examine whether the constraints
(\ref{1}) - (\ref{4}) admit solutions corresponding to $(4,4)$
supersymmetric WZNW sigma
models on the group manifolds from the list
given in \cite{Belg1}. Only for the simplest manifolds from this
list, namely $[U(1)]^4$ and $SU(2)\times U(1)$, the left and right
complex structures commute \cite{RSS} and only for the related
WZNW models there exists a description via twisted multiplets (in the
$q^{1,1}$ language, these models are described by the free action
(\ref{free}) and the action (\ref{wzwaction}), respectively). On
higher-dimensional manifolds which are not reduced to products of these
two, the left and
right structures do not commute. We conjecture that the associated
$(4,4)$ WZNW sigma models are described off shell by the
actions (\ref{haction}) with
proper potentials $h^{2,2\;[N,M]}$. The minimal number of the
superfield triples at which $h^{2,2\;[N,M]}$ exists, $n=2$, amounts to the
dimension 8 of the bosonic target. This precisely matches with the
dimension of the first nontrivial manifold from the aforementioned list,
that of the group $SU(3)$.

One more problem is to prove that the general action of the triples
$q^{1,1}, \omega^{1,-1}, \omega^{-1,1}$ in the analytic
$SU(2)\times SU(2)$
harmonic superspace indeed yields a most general
$(4,4)$ supersymmetric sigma model with torsion. Our starting point in
this paper was the analytic superfield $q^{1,1}$ which represents
a $(4,4)$ twisted multiplet.
But this is merely one type of $(4,4)$ twisted multiplet. There exist
other
types which display the same irreducible $(8+8)$ off-shell content, but
differ in the $SU(2)_L\times SU(2)_R$ assignment of component fields
(see, e.g., \cite{{GI},{Gates}}). For the time being it is unclear how to
simultaneously decribe all these types within the same $SU(2)\times SU(2)$
harmonic superspace. Perhaps, they can be related to each other by a
duality transformation (like all $N=2\;\;4D$ matter multiplets
are related to the ultimate analytic $q^{(+)}$ multiplet \cite{GIO}).
Alternatively, it may happen that for their consistent description
one will need to harmonize the whole $(4,4)$ supersymmetry
automorphism group  $SO(4)_L\times SO(4)_R$, i.e. to introduce two extra
sets of $SU(2)$ harmonic variables, and to consider appropriate analytic
superfields in this maximally extended $(4,4)$ harmonic superspace. The
relevant actions will be certainly more general than those discussed in
this paper. Clearly, in order to distinguish between
all these possibilities, one should understand in full the
geometry of the target space and various harmonic extensions of the
latter for general $(4,4)$ sigma models with torsion,
like this has been done for their hyper-K\"ahler counterparts
in \cite{GIOS1} and for $(4,0)$ sigma models in \cite{DKS}.

Finally, it would be interesting to find out possible implications of
the considered class of $(4,4)$ sigma models in the superstring theories
for which these sigma models could provide some consistent backgrounds.

\setcounter{equation}{0}
\def\thesection{ }
\section{Acknowledgements}

The major part of this
work has been accomplished during the author's visit to The Erwin
Schr\"odinger International Institute for Mathematical Physics in Vienna.
The author thanks Professors D. Alekseevsky and P. Michor for
their kind hospitality in ESI. He is grateful to Emery Sokatchev for
stimulating discussions of the geometry of $(4,4)$ models.
A partial support from the Russian Foundation of Fundamental Research,
grant 93-02-03821, and the International Science Foundation,
grants M9T000, M9T300 is also acknowledged.


\begin{thebibliography}{99}
\bibitem{4string} E. Kiritsis, C. Kounnas and D. L\"ust, Int. J. Mod.
Phys. {\bf A 9} (1994) 1361
\bibitem{GHR} S. J. Gates Jr., C. Hull and M. Ro\u{c}ek, Nucl. Phys.
{\bf B 248} (1984) 157
\bibitem{HoPa} P.S. Howe and G. Papadopoulos, Nucl. Phys. {\bf B 289}
(1987) 264; Class. Quantum Grav.
{\bf 5} (1988) 1647
\bibitem{AlFr} L. Alvarez-Gaum\'e and D.Z. Freedman, Commun. Math. Phys.
{\bf 80} (1981) 443; \\
J. Bagger and E. Witten, Nucl. Phys. {\bf B 222} (1983) 1
\bibitem{GIO} A. Galperin, E. Ivanov and V. Ogievetsky, Nucl. Phys.
{\bf B 282} (1987) 74
\bibitem{BGIO} J.A. Bagger, A.S. Galperin, E.A. Ivanov and
V.I. Ogievetsky, Nucl. Phys {\bf B 303} (1988) 522
\bibitem{GIOS1} A.S. Galperin, E.A. Ivanov, V.I. Ogievetsky and
E. Sokatchev, Ann. Phys. (N.Y.) {\bf 185} (1988) 22
\bibitem{GIOShk} A. Galperin, E. Ivanov, V. Ogievetsky and E. Sokatchev,
Commun. Math. Phys. {\bf 103} (1986) 515
\bibitem{GIOS} A. Galperin, E. Ivanov, V. Ogievetsky and E. Sokatchev,
JETP Lett. {\bf 40} (1984) 912
\bibitem{GIKOS} A. Galperin, E. Ivanov, S. Kalitzin, V. Ogievetsky
and E. Sokatchev, Class. Quant. Grav. {\bf 1} (1984) 469
\bibitem{DKS} F. Delduc, S. Kalitzin and E. Sokatchev, Class. Quantum
Grav. {\bf 7} (1990) 1567
\bibitem{DVal} F. Delduc and G. Valent, Class. Quantum Grav. {\bf 10}
(1993) 1201
\bibitem{ISu} E. Ivanov and A. Sutulin, Nucl. Phys. {\bf B 432} (1994)
246
\bibitem{BLR} T. Buscher, U. Lindstr\"om and M. Ro\u{c}ek, Phys. Lett.
{\bf B 202} (1988) 94
\bibitem{IK} E. A. Ivanov and S. O. Krivonos, J. Phys. A: Math.
and Gen. {\bf 17} (1984) L671
\bibitem{RSS} M. Ro\u{c}ek, K. Schoutens and A. Sevrin, Phys. Lett.
{\bf B 265} (1991) 303
\bibitem{IR} U. Lindstr\"om, I.T. Ivanov and M. Ro\u{c}ek, Phys. Lett.
{\bf B 328} (1994) 49
\bibitem{DS} F. Delduc and E. Sokatchev, Int. J. Mod. Phys. {\bf B 8}
(1994) 3725
\bibitem{Zum} B. Zumino. Phys. Lett. {\bf B 87} (1979) 203
\bibitem{nonlYM} N. Ikeda, Ann. Phys. (N.Y.) {\bf 235} (1994) 435; \\
P. Schaller and T. Strobl, Mod. Phys. Lett. {\bf A 9} (1994) 3129
\bibitem{Belg1}
Ph. Spindel, A. Sevrin, W. Troost and A. Van Proeyen,
Phys. Lett. {\bf B 206} (1988) 71; Nucl. Phys. {\bf B 308} (1988) 662
\bibitem{GI} O. Gorovoy and E. Ivanov, Nucl. Phys. {\bf B 381}
(1992) 394
\bibitem{Gates} S. James Gates, Jr., Phys. Lett. {\bf B 338} (1994) 31

\end{thebibliography}
\end{document}